\documentclass[12pt]{article}

\usepackage{geometry}
\usepackage{amssymb,amsmath,amsfonts,eurosym,ulem,graphicx,caption,color,setspace,sectsty,comment,natbib,pdflscape, array}
\usepackage{hyperref}
\hypersetup{
colorlinks=true,
citecolor=black,
linkcolor=black,
filecolor=black,      
urlcolor=black,
} 

\usepackage{tikz-qtree}
\usepackage{bbold}
\usepackage[bottom]{footmisc}
\usepackage{algorithm, algpseudocode}
\usepackage{tabularx}
\usepackage{caption}
\usepackage{subcaption}
\usepackage{hyperref}
\usepackage{float}
\usepackage{graphicx}
\usepackage{booktabs}
\usepackage[flushleft]{threeparttable}
\usepackage{bm}
\usepackage{setspace}
\usepackage[color=orange!20]{todonotes} \presetkeys{todonotes}{inline}{}

\newcommand{\bX}{\mathbf{X}}

\DeclareMathOperator*{\argmin}{arg\,min}

\usepackage{scalerel,stackengine}
\usepackage{color}
\allowdisplaybreaks
\stackMath
\newcommand\reallywidehat[1]{%
	\savestack{\tmpbox}{\stretchto{%
			\scaleto{%
				\scalerel*[\widthof{\ensuremath{#1}}]{\kern-.6pt\bigwedge\kern-.6pt}%
				{\rule[-\textheight/2]{1ex}{\textheight}}
			}{\textheight}%
		}{0.5ex}}%
	\stackon[1pt]{#1}{\tmpbox}%
}

\usepackage{scalerel}
\usepackage{stackengine,wasysym}

\makeatletter
\newcommand*{\indep}{%
	\mathbin{%
		\mathpalette{\@indep}{}%
	}%
}
\newcommand*{\nindep}{%
	\mathbin{
		\mathpalette{\@indep}{\not}
	}%
}
\newcommand*{\@indep}[2]{%
	\sbox0{$#1\perp\m@th$}
	\sbox2{$#1=$}
	\sbox4{$#1\vcenter{}$}
	\rlap{\copy0}
	\dimen@=\dimexpr\ht2-\ht4-.2pt\relax
	\kern\dimen@
	{#2}%
	\kern\dimen@
	\copy0 
} 
\makeatother

\newtheorem{assumption}{Assumption}

\newcolumntype{L}[1]{>{\raggedright\let\newline\\arraybackslash\hspace{0pt}}m{#1}}
\newcolumntype{C}[1]{>{\centering\let\newline\\arraybackslash\hspace{0pt}}m{#1}}
\newcolumntype{R}[1]{>{\raggedleft\let\newline\\arraybackslash\hspace{0pt}}m{#1}}

\begin{document}

\begin{titlepage}

\title{\Large Transporting Predictions via Double Machine Learning:\\ Predicting Partially Unobserved Students'  Outcomes}
\author{\normalsize Falco J. Bargagli-Stoffi\thanks{\scriptsize Corresponding author. Mail to: \href{mailto: falco@ucla.edu}{falco@ucla.edu}. University of California, Los Angeles (UCLA), Los Angeles, CA, United States.} \and
\normalsize Emma Landry \thanks{\scriptsize University of California, Los Angeles (UCLA), Los Angeles, CA, United States.} \and
\normalsize Kevin P. Josey\thanks{\scriptsize Colorado School of Public Health, Aurora, CO, United States.} \and 
\normalsize Kenneth De Beckker\thanks{\scriptsize Open Universiteit, Heerlen, The Netherlands \& KU Leuven, Brussels, Belgium.} \and \normalsize Joana Elisa Maldonado\thanks{\scriptsize European Commission, Brussels, Belgium. The information and views set out in this article are those of the author and do not necessarily reflect the official opinion of the European Commission.} \and \normalsize Kristof De Witte\thanks{\scriptsize KU Leuven, Leuven, Belgium \& UNU-Merit, Maastricht University, Maastricht, The Netherlands.} }
\date{}
\maketitle
\vspace{-1.25cm}
\begin{abstract}
\noindent{Educational policymakers often lack data on student outcomes where standardized tests were not administered. Machine learning can predict unobserved outcomes in target populations using source population data. However, covariate distribution differences between populations reduce model transportability, potentially decreasing predictive accuracy and introducing bias. We propose using double machine learning for covariate-shift weighted models. First, we estimate overlap scores---the probability an observation belongs to the source dataset given covariates. Second, balancing weights, defined as density ratios of target-to-source membership probabilities, reweight individual observations' contributions to the loss function in target outcome prediction models. This downweights source observations less similar to the target population, allowing predictions to rely more on observations with greater overlap. Consequently, predictions become more transportable under covariate shift. We illustrate this framework using student standardized financial literacy scores (FLS) data. Using Bayesian Additive Regression Trees (BART), we predict missing FLS. We find minimal predictive performance differences between weighted and unweighted models, suggesting limited covariate shift in our setting. Nonetheless, our approach provides a principled framework for addressing covariate shift and is broadly applicable to predictive modeling in social and health sciences, where source-target population differences are common.}\\ 

\vspace{-0.50cm}\\
\end{abstract}
\setcounter{page}{0}
\thispagestyle{empty}
\end{titlepage}

\doublespacing

\section{Introduction} \label{sec:introduction}

Machine learning has become increasingly prominent in social and health sciences as it offers powerful tools for prediction, causal inference, theory testing, and the development of new data sources \citep{ludwig2024machine}. This rise has been mainly driven by the staggering performance of machine learning in predictive tasks. However, using off-the-shelf machine learning methodologies to perform predictions requires careful consideration. A key challenge arises when the training data differs substantially from the data used for prediction. We propose a novel methodology to detect regions of the predictors' space in the out-of-sample set where the algorithm relies on extrapolation.

To build a predictive model, covariate and outcome data are used from a source population. However, there is no guarantee that observations in the source population are sampled from the same population as the target. For instance, predicting financial literacy scores for students in one region based on data from another region with different socioeconomic characteristics may lead to poor transportability of the predictions, reducing the performance quality and introducing bias. 

The scenario where the covariate distributions vary between source and target populations, but the conditional distribution of the outcome is exchangeable, is referred to as \textit{covariate shift} \citep{shimodaira2000improving}. Approaches aiming to tailor prediction models to mitigate effects of such shifts are referred to as \textit{domain adaptation} in the statistical and computer science literature \citep{sugiyama2007covariate, huang2006correcting, bickel2009discriminative, baktashmotlagh2014domain}, and fall under the larger umbrella of transfer learning. \cite{kouw2019review} provide a comprehensive overview of domain adaptation and describe other types of distributional shifts that may occur in empirical data.

In addition to the machine learning literature, we note parallels with the concepts of \textit{generalizability} and \textit{transportability} in causal inference \citep{degtiar2023review}. In that context, re-weighting procedures are introduced to generalize experimental results from a source to a target study by adjusting for differences in baseline covariate distributions \citep{dahabreh2019generalizing, egami2021covariate}. Our framework connects to recent developments in prediction under covariate shift. In particular, \cite{yang2024doubly} formalize the connection between covariate shift and the missing at random (MAR) assumption, distinguishing between \textit{explainable covariate shift} \citep{tibshirani2019conformal}---where the shift can be fully accounted for by observed covariates---and settings where unobserved factors drive the distributional differences. Our framework operates under the explainable covariate shift assumption, which is equivalent to assuming that the outcome is MAR conditional on observed covariates. While \cite{yang2024doubly} develop a conformal prediction framework for constructing prediction sets with coverage guarantees, we focus on point prediction using importance-weighted machine learning within a double machine learning framework. Both approaches share the core insight of estimating the likelihood ratio to reweight observations from the source population.

Methodologies to address covariate shift typically rely on reweighting of the source dataset to align the source feature space with the target. We follow a similar approach to \cite{steingrimsson2023transporting} for weight estimation, which we refer to as \textit{double machine learning}. In a first stage, we estimate the \textit{overlap score}, which is the probability of a given observation belonging to the source dataset conditional on its covariates. \textit{Balancing weights} are then calculated as the density ratio of the probability of belonging to the target to the probability of belonging to the source \citep{ben2021balancing}. These weights are used to scale individual observations' loss or likelihood contributions in the prediction model for the target outcome of interest. 

This approach is general enough to be applied to both Bayesian methodologies \cite[i.e., the Bayesian Additive Regression Tree (BART) method introduced by][]{chipman2010bart} and frequentist techniques \citep[i.e., the random forest methodology by][]{breiman2001random}. Its main goal is to downweight source observations that are considered different from the target population, and let the predictions rely more heavily on observations with a greater overlap. This targeted weighting improves the relevance of source data for prediction, reduces bias from covariate shift, and enhances model transportability. It thereby addresses a key limitation in much of the existing literature, which often assumes strong overlap or fails to adequately account for differences in covariate distributions between source and target populations.

As a motivating application, we are the first to introduce supervised machine learning in the financial literacy literature. First, we use machine learning to predict financial literacy scores (FLS) for students in a region of Belgium where these scores are not observed. We train our model using OECD PISA data \citep{OECD2017a} for the Flemish region of Belgium where the FLS are observed. Then, we predict the missing FLS of students in the Walloon region of Belgium. We use an extensive set of student and school-level variables as predictors in our machine learning model. 

Second, we compare predictions obtained under the standard prediction model to those obtained while addressing covariate shift through reweighting. Covariate shift remains largely ignored in the education literature, with the exception of \cite{luzan2024evaluation} who employ kernel mean matching (KMM) to aid predictions of student attrition in higher education. While not explicitly focusing on covariate shift, the work of \cite{ren2024alignment} presents a relevant approach to the analysis of the PISA data, where the authors adopt a potential outcomes approach to determine transportability across countries in the study. 

Third, we use machine learning to identify the characteristics of students with outlying predicted test scores. In our case, the main focus is on students with lower predicted FLS. Indeed, machine learning can help researchers and policymakers to understand the profiles of students at risk of underperformance. These insights could inform the design of targeted interventions aimed at improving the outcomes of low-performing students. 

Our main results can be summarized as follows. The proposed model depicts a strong performance achieving an adjusted $R^2$ of roughly 73\% with relatively small root mean squared error (RMSE) and mean absolute error (MAE).\footnote{This performance measures result from 10-folds cross-validation, under a standard unweighted model.} Moreover, BART, allows us to get draws from the posterior distribution of the predicted observations and construct credible intervals for the unit-level predictions. Hence, we are able to detect the predictions that fall outside the 95\% credible intervals for the mean posterior predicted values. We use this information to identify a set of outlier predictions. 

On the one hand, this analysis shows that the predictive  probability  of  having  a  low  financial  literacy  score  is  the  largest for  students  with  lower  scores for  reading  and math  in  the  PISA  test. On the other hand, the students' background plays a critical role: students with the lowest predicted FLS are often individuals from families where the school language is not spoken at home, and the parents have a poor educational background. Moreover, we find that our proposed methodology for addressing covariate shift does not improve the quality of predictions compared to standard BART modeling, hinting that covariate shift may not be a severe issue in the prediction of FLS for students in Belgium. The BART model appears robust enough to transport predictions from the source to the target population. 

The remainder of the paper is organized as follows. In Section \ref{sec:covariateshift}, we introduce the machine learning methodologies used and developed in this paper. In Section \ref{sec:simulations}, we evaluate the performance of the proposed methodology via Monte Carlo simulations.  Section \ref{sec:data} describes the PISA data used to illustrate the proposed methodology. In Section \ref{sec:results}, we discuss the results obtained from the application on the PISA data. Section \ref{sec:discussion} concludes with a discussion of the results.\footnote{The \texttt{R} and \texttt{Stata} codes used for the analysis, together with the data and the functions for the machine learning analysis, will be made publicly available on the GitHub page of the corresponding author (\href{https://github.com/fbargaglistoffi/transport_predictions}{\texttt{https://github.com/fbargaglistoffi/transport\_predictions}}).}

\section{Methodology: Transporting Predictions via Double Machine Learning}\label{sec:covariateshift}

This Section introduces the methodologies for covariate shift that extend the usage of machine learning algorithms to address transportability of predictions and the detection of observations with outlying predicted values.

\subsection{Problem Setup and Assumptions}

Let us introduce an indicator $S \in \{0,1\}$ that identifies whether the data are sampled from a \textit{source} or \textit{target} population, with $S= 1$ for the source population and $S= 0$ for the target population. For both the source and target population, we observe the set of predictors $\bX$, but the vector of outcomes is observed just for the source population.

Our goal is to construct a predictive model using a random sample of the source population $\mathcal{S} = \{ (y_i, \mathbf x_i): i \in \{j: S_j = 1\}\}$, that optimize predictions for the target population on a random sample $\mathcal{S}^c = \{ (\mathbf x_i) : i \in \{j: S_j = 0\}\}$.  For a generic machine learning algorithm, the objective is to estimate the prediction function $f(\mathbf{X})$ using a loss function $\ell(f(\mathbf{X}), Y)$  such that
\begin{equation}
\hat{f}(\mathbf{x}) = \argmin_{f \in \mathcal{F}} \int_{\mathcal{X}} \int_{\mathcal{Y}} \: \ell(f(\mathbf{x}), y) \: f(y \mid \mathbf{x})\: f(\mathbf{x}) dy dx.
\end{equation} 
However, this objective assumes the population where training and test data are sampled is the same as the population where the model will be implemented. This assumption is often violated whenever a prediction model is transported to a different population. It is important to distinguish two related but distinct inferential targets. First, under Assumption~\ref{assump:exch}, the conditional expectation function $f(\mathbf{x}) = \mathbb{E}[Y \mid \mathbf{X} = \mathbf{x}]$ is invariant across populations and can be estimated consistently using source data alone. Second, the marginal mean in the target population, $\mathbb{E}_{S=0}[Y] = \int f(\mathbf{x}) \, dF_{\mathbf{X} \mid S=0}(\mathbf{x})$, requires integration over the target covariate distribution. Importance weighting addresses the latter by reweighting the source sample to approximate the target distribution. For individual-level predictions $\hat{f}(\mathbf{x}_i)$ where $\mathbf{x}_i$ is observed in the target, the conditional expectation function suffices and weighting offers no direct benefit for point prediction, a consideration relevant to interpreting our empirical findings.

Under covariate shift, the input, or covariate, distributions are assumed to differ between the source and target populations, such that $f(\mathbf{x} \mid S = 1) \ne f(\mathbf{x} \mid S = 0)$. To ensure model identifiability, we make the following two assumptions, standard in the prediction transportability literature \citep{steingrimsson2023transporting}.

\begin{assumption}[Conditional sample exchangeability]\label{assump:exch}
    For any $\mathbf X$ such that $f(\mathbf X = \mathbf{x}, S = 0) \ne 0$, $f(Y \mid \mathbf{X}=\mathbf{x}, S = 1) = f(Y \mid \mathbf X = \mathbf{x}, S = 0)$. 
\end{assumption}

\vspace{0.25cm}

\begin{assumption}[Positivity]\label{assump:posi}
    For any $\mathbf X$ such that $f(\mathbf X = \mathbf{x} , S = 0) \ne 0$, $Pr(S = 1 \mid \mathbf X = \mathbf{x}) > 0.$
\end{assumption}

The first assumption implies that the joint distribution of the outcome $Y$ and covariates $\mathbf{X}$ differs only in terms of the covariate distribution. Following \cite{yang2024doubly}, we note that the covariate shift setting considered here corresponds to explainable covariate shift \citep{tibshirani2019conformal}, where the distributional differences between source and target populations can be fully characterized by observed covariates $\mathbf{X}$. Formally, this is equivalent to assuming that the outcome $Y$ is missing at random (MAR) in the target population, i.e., $Y \perp S \mid \mathbf{X}$.The second assumption ensures that any covariate pattern that appears in the target population can appear in the source population with positive probability. 

Under these assumptions, we can consistently estimate predictions for the target population by appropriately reweighting the source data. The following subsections develop an \textit{importance weighting approach} within a \textit{double machine learning framework} suited for this task.

\subsection{Probabilistic Overlap Scores and Importance Weighting for Covariate Shift}

To obtain predictions optimized for the target population, we must account for the difference in covariate distributions between source and target. The key insight is that we can reweight the source observations so that, after reweighting, they resemble a sample from the target distribution.

To do so, we can design a weighted loss function that is adjusted by an estimate of the density ratio $\frac{f(\mathbf{X} \mid S = 0)}{f(\mathbf{X} \mid S = 1)}$ which is referred to as a \textit{balancing weight}, \textit{likelihood ratio} or \textit{importance weight} in the covariate shift literature \citep{yang2024doubly, shimodaira2000improving}. This density ratio quantifies how much more (or less) likely a given covariate pattern is to appear in the target versus the source population.
 This strategy is often referred to as \textit{data importance-weighting} and a wide range of approaches have been explored to obtain estimates of this ratio, many of which focus on optimizing a measure of dissimilarity using method-of-moment type estimators (see \cite{kouw2019review} for a comprehensive overview). Here, we avoid parametric assumptions, instead adopting a probabilistic classification approach. 

Specifically, we follow the methodology of \cite{steingrimsson2023transporting} who proceed with the inverse-odds of being from the source population instead of the density ratio, since
\[
\frac{f(\mathbf{X} \mid S = 0)}{f(\mathbf{X} \mid S = 1)} \propto \frac{Pr(S = 0 \mid \mathbf{X})}{Pr(S = 1 \mid \mathbf{X})}, \quad \text{by Bayes's rule. } 
\]
This yields the weighted loss optimization objective to solve \begin{equation}\label{eq:objective}
\begin{split} 
f^*(\mathbf{x}) = \argmin_{f \in \mathcal{F}} 
&\int_{\mathcal{X}} \int_{\mathcal{Y}} \frac{Pr(S = 0 \mid \mathbf{x})}{Pr(S = 1 \mid \mathbf{x})} \: \ell(f(\mathbf{x}), y) \: f(y \mid \mathbf{x}) \: f(\mathbf{x} \mid S = 1) \: dydx \\ 
\propto \argmin_{f \in \mathcal{F}} 
&\int_{\mathcal{X}} \int_{\mathcal{Y}} \frac{f(\mathbf{x} \mid S = 0)}{f(\mathbf{x} \mid S = 1)} \: \ell(f(\mathbf{x}), y) \: f(y \mid \mathbf{x}) \: f(\mathbf{x} \mid S=1) \: dydx\\
= \argmin_{f \in \mathcal{F}}
&\int_{\mathcal{X}} \int_{\mathcal{Y}} \: \ell(f(\mathbf{x}), y) \: f(y \mid \mathbf{x}) \: f(\mathbf{x} \mid S=0) \: dydx\end{split}
\end{equation}
which is the desired loss for a target population described by the covariate distribution in the sample $\mathcal{S}^c$. 

Since the ratio $\frac{Pr(S = 0 \mid \mathbf{X})}{Pr(S = 1 \mid \mathbf{X})}$ is unknown, we can estimate $\hat{Pr}(S = 1 \mid \mathbf{X})$, which we will refer to as the \textit{overlap score}, using machine learning.  We can then obtain $\hat{Pr}(S = 0 \mid \mathbf{X}) = 1 - \hat{Pr}(S = 1 \mid \mathbf{X})$. 

The estimated overlap scores can then be incorporated into the empirical version of the objective function in (\ref{eq:objective}) to obtain 
\begin{equation}\label{eq:new_obj}
\hat{f}^* = \argmin_{f \in \mathcal{F}} \sum_{i \in \mathcal{S}} \frac{\hat{Pr}(S_i = 0 \mid \mathbf{X}_i)}{\hat{Pr}(S_i = 1 \mid \mathbf{X}_i)} \: \ell(f(\mathbf{X}_i), Y_i). 
\end{equation}
This weighted objective function upweights source observations that are more representative of the target population while downweighting those that are less representative, effectively rebalancing the training data to better match the target distribution.

\subsection{Double Machine Learning for Transported Predictions}\label{subsec:dml}

The transported prediction problem requires estimating two unknown quantities: (i) the overlap scores $Pr(S=1 \mid \mathbf{X})$ that define the importance weights, and (ii) the prediction function $f(\mathbf{X})$ that minimizes weighted loss on the target population. We frame this as a double machine learning problem, where the overlap 
scores serve as a nuisance function estimated in a first stage, and the prediction model is estimated in a second stage using the fitted weights.

This two-stage structure is analogous to propensity score methods in causal inference \citep{rosenbaum1983central}, where treatment assignment probabilities are first estimated, then used to reweight or match observations. Here, the overlap score plays a similar role: it estimates the probability of population membership, which determines how much each source observation should contribute 
to the transported prediction model. 

A critical motivation for this approach is to address potential bias arising from distributional differences between populations. Consider a prediction function $\hat{f}(\mathbf{X})$ fitted on source data. While prediction errors may satisfy $\big(Y - \hat{f}(\mathbf{X})\big) \perp \mathbf{X} \mid S = 1$ in the source, this does not guarantee $\big(Y - \hat{f}(\mathbf{X})\big) \perp \mathbf{X} \mid S = 0$ in the target, potentially yielding biased transported predictions. The conditional exchangeability assumption (Assumption~\ref{assump:exch}) ensures that for the true prediction function $f^*(\mathbf{X})$:
\[ \big(Y - f^*(\mathbf{X})\big) \perp \mathbf{X} \mid S = 1 \implies 
\big(Y - f^*(\mathbf{X})\big) \perp \mathbf{X} \mid S = 0. \]
However, this implication may fail for $\hat{f}(\mathbf{X})$ when the model is misspecified. Using flexible machine learning methods for both stages---overlap score estimation and prediction---mitigates this concern by reducing dependence on correct functional form specification.

Under Assumptions~\ref{assump:exch}--\ref{assump:posi}, the double machine learning approach yields consistent transported predictions. If the overlap score estimator is consistent, the importance-weighted prediction function converges to the oracle that minimizes target-population risk:
\[
\hat{f}^* \xrightarrow{p} \argmin_{f \in \mathcal{F}} 
\mathbb{E}_{S=0}[\ell(f(\mathbf{X}), Y)].
\]
This follows because importance weighting, under correct specification of the density ratio, transforms the source distribution to match the target \citep{shimodaira2000improving, sugiyama2007covariate}. The positivity assumption ensures bounded weights, preventing infinite variance.

We maintain the weighted double machine learning approach in our analysis for three reasons: (1) it provides a theoretically principled baseline that is consistent under correct specification and sufficient overlap;\footnote{When covariate overlap is insufficient, importance weighting may perform poorly, and practitioners should consider restricting predictions to regions of adequate overlap.} (2) it demonstrates the practical challenges of importance weighting in high-dimensional settings; and (3) it motivates our analysis of when simpler unweighted methods may suffice.

\subsection{Implementation}\label{subsec:implementation}

We implement the DML framework for transported predictions in two stages. In the first stage, we estimate overlap scores using a classification model for population membership. In the second stage, these scores are used to construct importance weights that are incorporated into the prediction model.

\textit{Overlap Score Estimation.} The overlap scores $Pr(S=1 \mid \mathbf{X})$ can be estimated using generalized linear models for binary outcomes, such as logistic or probit regression, or flexible machine learning methods for classification, such as classification trees \citep{friedman1984classification} and random forests \citep{breiman2001random}. There are also alternative 
approaches that estimate the density ratio $\frac{Pr(S=0 \mid \mathbf{X})}{Pr(S=1 \mid \mathbf{X})}$ directly using method-of-moment type estimators, often referred to as \textit{balancing weights} \citep{josey2021framework}. We note that under the non-nested sampling design, where samples from source and target populations are obtained independently, the inverse-odds weights are non-identifiable from the data as sampling fractions from the superpopulations are unavailable. However, \cite{steingrimsson2023transporting} demonstrate that the population inverse-odds weights are proportional to those in the training set. 

\textit{Assessing Covariate Overlap.} The estimated overlap scores can also be used to assess \textit{covariate overlap}---the degree to which the covariate distributions of source and target populations share common support. Poor overlap occurs when the target population contains covariate patterns that are rare or absent in the source, leading to extreme weights that increase prediction variance. In practice, overlap can be assessed by examining the distribution of estimated scores: values concentrated near 0 or 1 indicate regions of poor overlap.

\textit{Weighted Prediction Model.} Given estimated overlap scores, the importance weights $\hat{\omega}_i = \frac{1 - \hat{Pr}(S_i=1 \mid \mathbf{X}_i)}{\hat{Pr}(S_i=1 \mid \mathbf{X}_i)}$ 
are incorporated into the empirical objective:
\begin{equation}\label{eq:empirical_obj}
\hat{f}^* = \argmin_{f \in \mathcal{F}} \sum_{i \in \mathcal{S}} \hat{\omega}_i 
\, \ell(f(\mathbf{X}_i), Y_i).
\end{equation}
This weighted objective upweights source observations more representative of the target population while downweighting those less representative, effectively rebalancing the training data to match the target distribution.

\subsection{BART for Prediction}\label{subsec:bart}

The importance-weighted prediction framework developed above is agnostic to the choice of prediction model $\hat{f}(\mathbf{X})$. Any supervised learning method that accepts observation weights can be used. We focus on Bayesian Additive Regression Trees (BART) \citep{chipman2010bart} for three reasons: (1) BART consistently achieves strong predictive performance across diverse settings without extensive tuning both in prediction tasks \citep{murray2021log, linero2018bayesian1, linero2018bayesian2, hernandez2018bayesian, bargagli2024machine} and in causal inference tasks \citep{hill2011bayesian, logan2019decision, nethery2019estimating, bargagli2022heterogeneous, wang2024local, englert2025estimating}; (2) its Bayesian formulation naturally accommodates observation weighting through likelihood reweighting, as we describe below; and (3) the posterior predictive distribution provides principled uncertainty quantification, enabling the outlier detection analysis in Section~\ref{subsec:outliers}.

BART finds its foundation in the CART algorithm \citep{friedman1984classification}. CART is a widely used algorithm for the construction of trees where each node is split into only two branches (i.e., binary trees). The accuracy of the predictions of binary trees can be dramatically improved by iteratively constructing the trees. BART is a sum-of-trees ensemble algorithm, and its estimation approach relies on a fully Bayesian probability model \citep{Kapelner2016}. 

Let us define with $Y$ the outcome vector, with $y_i$ the outcome for a generic unit $i$, and with $\bX$ the $N \times P$ matrix of covariates or \textit{predictors} (where $P$ is the number of predictors), with $X_i$ the $P$-dimensional vector of predictors for $i$ (with $i=1,..., N)$.
Then, the BART model can be expressed as:
\begin{equation}\label{tree}  \centering 
    Y =  \mathcal{T}_1(\bX; \mathcal{D}_1, \mathcal{M}_1) + ... + \mathcal{T}_J(\bX; \mathcal{D}_j, \mathcal{M}_j) + \epsilon, \:\:\:\:\:\:\:\:\: \epsilon_i \sim \mathcal{N} (0, \sigma^2),
\end{equation}
where the $J$ distinct binary trees are denoted by $\mathcal{T}(\bX; \mathcal{D}_j, \mathcal{M}_j)$. $\mathcal{T}$ is a function that sorts each unit into one of the sets of $m_j$ terminal nodes, associated with mean parameters $\mathcal{M}_j = \{\mu_1, ..., \mu_{m_j}\}$, based on a set of decision rules, $\mathcal{D}_j$.
$\epsilon$ is an error term and is typically assumed to be independent and identically normally distributed when the outcome is continuous \citep{chipman2010bart}.   
The Bayesian component of the algorithm is incorporated in a set of three different priors on: (i) the structure of the trees, $\mathcal{D}_j$ (this prior is aimed at limiting the complexity of any single tree $\mathcal{T}$ and works as a regularization device); (ii) the distribution of the outcome in the nodes, $\mathcal{M}_j$ (this prior is aimed at shrinking the node predictions towards the center of the distribution of the response variable $Y$);  (iii) the error variance $\sigma^2$ (which bounds away $\sigma^2$ from very small values that would lead the algorithm to overfit the training data).\footnote{The choice of the priors and the derivation of the posterior distributions is discussed in depth by \cite{chipman2010bart} and \cite{Kapelner2016}. Namely, (i) the prior on the probability that a node will split at depth $k$ is $\beta(1+k)^{-\eta}$ where $\beta \in (0,1), \eta \in [0, \infty)$ (these hyper-parameters are generally chosen to be $\eta=2$ and $\beta = 0.95$); (ii) the prior on the probability distribution in the nodes is a normal distribution with zero mean: $\mathcal{N}(0, \sigma^2_q)$ where $\sigma_q = \sigma_0/\sqrt{q}$ and $\sigma_0$ can be used to calibrate the plausible range of the regression function; (iii) the prior on the error variance is $\sigma^2 \sim InvGamma(v/2, v\lambda/2)$ where $\lambda$ is determined from the data in a way that the BART will improve 90\% of the times the RMSE of an OLS model.} The aim of these priors is to ``regularize" the algorithm, preventing single trees from dominating the overall fit of the model \citep{Kapelner2016}. 
 
These Bayesian tools give researchers the possibility to mitigate the overfitting problem of random forests \citep{breiman2001random} and to tune the algorithm with prior knowledge. Moreover, BART has shown a consistently strong performance under ``default" model specifications \citep{chipman2010bart}. This is a highly valuable characteristic of BART as it reduces its dependence on the choice of parameters done by the researcher as well as the computational time and costs related to cross-validation.

We note that, in the context of Bayesian statistics, the likelihood $f(Y \mid \theta)$ for a model governed by parameters $\theta$ always defines a loss function $\ell (Y; \theta) = - \log f(Y \mid \theta)$ \citep{good1952rational}. Each observation's likelihood can then be raised to the power by a balancing weight; this reweighted density has been demonstrated to integrate to one for the exponential family of distributions \citep{bernardo2009bayesian, wang2017robust}. The subsequent analyses rely on predictions using BART, for which the model likelihood is Gaussian. Under this setting, weighting individual observations is in fact equivalent to scaling individual variances as $\sigma^2_i = \sigma^2 / \omega_i$, where $\omega_i$ are the individual weights given by the density ratio $\frac{\hat{Pr}(S_i = 0 \mid \mathbf{X}_i)}{\hat{Pr}(S_i = 1 \mid \mathbf{X}_i)}$. 

\subsection{Predictive Outliers Detection}\label{subsec:outliers}

From the two-step double machine learning procedure introduced in the previous section, we obtain draws from the posterior predictive distribution of $Y$. Then, we can use standard techniques employed for outlier detection, such as the ones proposed by \cite{miller1991reaction} (i.e., $K$ standard deviation from the mean of the distribution) and \cite{leys2013detecting} (i.e., $K$ absolute deviations from the median), to detect the predictions with values that are consistently further away from the mean or the median of the predicted distribution, respectively. The nice feature of such techniques is that they can be manually tuned in order to include units with more or less extreme predicted values.\footnote{In case of multivariate predictions, one could implement data-driven methodologies for outlier detection, such as the \textit{Isolation Forest} \citep{liu2012isolation}.}

These analyses can be relevant for practitioners and policy-makers to spot observations with more or less extreme predicted outcome values. Indeed, it could be the case that policy-makers are interested in targeting a specific subpopulation, based on the values of an unobserved, yet predictable, outcome (e.g., provide additional learning material to more vulnerable students in a region where financial literacy is not observed). Moreover, information on the factors related to these ``high or low levels" of the predicted outcome are not only useful to identify potential subgroups but also to reveal which factors are associated with specific levels or values of the outcome.

\section{Simulation Study} \label{sec:simulations}

\subsection{Simulation Set-up}

Prior to proceeding with the analyses of the financial literacy scores, we illustrate the covariate shift methodology introduced in Section~\ref{sec:covariateshift} under a controlled covariate shift simulation framework. Specifically, our goal is to compare the predictive performance of a standard machine learning model to that of its importance weighted counterpart to identify the scenarios in which weighting helps, and those in which it does not. 

For our data generating process, we assume there are four covariates ($X_1, X_2, X_3, X_4)$, whose marginal distributions differ between a source and target population. We let $X_p^\mathcal{S}$ and $X_p^\mathcal{T}$ denote the $p^\text{th}$ covariate in the source and target, respectively. The covariates are then generated from the following mechanism: $X_1^\mathcal{S} \sim \mathcal{N}(0, 3^2)$ and $X_1^\mathcal{T} \sim \mathcal N (0, 9^2)$; $X_2^\mathcal{S} \sim \mathcal{N}(5, 3^2)$ and $X_2^\mathcal{T} \sim \mathcal N (9, 3^2)$; 
$X_3^\mathcal{S} \sim \mathcal{N}(-2, 4^2)$ and $X_3^\mathcal{T} \sim \mathcal N (0, 1^2)$; 
$X_4^\mathcal{S} \sim \text{Binomial}(4, 0.3)$ and $X_4^\mathcal{T} \sim \text{Binomial}(4, 0.7)$.

The target population is assumed to be of size of $m$, where all covariates are generating following the associated mechanism. On the other hand, the source population has size $n_1 + n_2$, where $n_1$ is the number of observations whose covariates are generated using the source mechanism, and $n_2$ the number of observations whose covariates are generated using the target mechanism. Therefore, the smaller the value of $n_1$, the lesser the overlap between the source and target populations. 

The outcome function $f$ is defined as $f(x_1, x_2, x_3, x_4) = 3 + \sin(x_1) + \log(1 + x_2^2) + 0.5 \cdot x_3 \cdot x_4 + x_2 \mathbb{I}\{x_4 > 2\}$. We evaluate the outcome function on $m$ sets of covariates from the target set, contaminate the values with $\mathcal N(0,1)$ noise, and train a BART model on this data. Our outcome data $Y^\mathcal S$ and $Y^\mathcal T$ is then obtained as the values predicted by BART, contaminated by $\mathcal N(0,2)$ noise. The data generating mechanism is therefore based on the target data, but in order to replicate the covariate shift set-up, analysis is conducted on the source data under the assumption that target labels would in practice be unobserved. This set-up also ensures that the key assumption of conditional sample exchangeability is satisfied.

\subsection{Results}

We conduct a Monte Carlo study comparing three scenarios. We assume that the size of the target population is $m= 150$, and consider three scenarios. Namely, in Scenario 1 we have $n_1 = n_2 = 100$; in Scenario 2 $n_1 = 150, n_2 =50$ and in Scenario 3 $n_1 = 200, n_2 = 0$. Scenario 1 corresponds to a case of high overlap between source and target population (small covariate shift); Scenario 2 depicts a case of moderate overlap (moderate covariate shift); Scenario 3 corresponds to the case of low overlap (large covariate shift).

In each scenario, we simulate 40 random datasets and train an unweighted BART model, a Random Forest model, and an importance weighted BART model on the source dataset. We note that an importance weighted Random Forest is not considered as it is not a likelihood-based approach. We then obtain predictions on the target set and compute the RMSE across the $m$ observations. The comparison of the marginal covariate distributions between source and target for the three scenarios is provided for reference in Appendix Figure~\ref{fig:cov_distributions}.

\begin{figure}[ht!]
    \centering
        \includegraphics[width=0.99\textwidth]{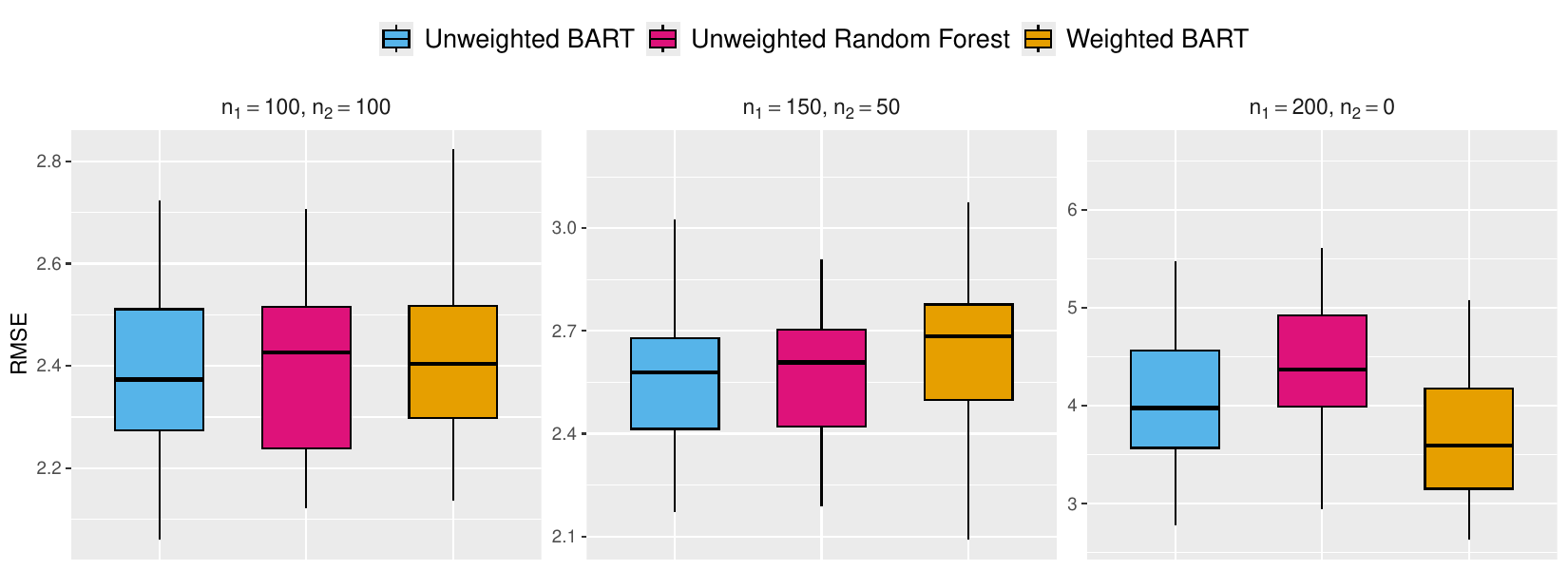} 
    \caption{Boxplot of RMSE values across 40 simulated datasets. Each plot corresponds to one of the three data generation scenarios considered. In blue are the results with the standard BART model; in pink, those with a Random Forest model; and in orange, those obtained with weighted BART model.}
    \label{fig:sim_results}
\end{figure}

The results of our simulation study are illustrated in Figure~\ref{fig:sim_results}. For each of the three considered scenarios (ranging from most to least overlap between the source and target populations), we obtain the RMSE across 40 Monte Carlo simulations, and represent them in a boxplot. We note that for all three scenarios, the Random Forest displays performance comparable to unweighted BART, with slightly greater discrepancy for Scenario 3. We note that the generative model for the data is based on a BART construction, but given the non-parametric nature of the algorithms, we do not expect it to be a significant contributor differences in performance.

Focusing on the comparison between unweighted and weighted BART, we find that in Scenarios 1 and 2, the former model outperforms the latter in terms of average error across observations in the target dataset. We note that these are the cases of greater overlap in (marginal) covariate distributions between source and target. These results seem to indicate that when covariate shift is minimal, importance weighting does not help the model's predictive ability on target data, but instead induces extra noise that hurts performance. On the other hand, in the case where $n_2 =0$, that is strongest covariate shift, the weighted model displays lower errors than standard BART. The RMSE values are overall higher, which is explained by the fact that the machine learning model is trained on data that differs more significantly from the one on which predictions are made and evaluated. These results showcase the importance of re-weighting observations when covariate shift is apparent. 

The simulation study demonstrates that the use of an importance weighted model improves performance under pronounced covariate shift. However, if covariate shift's impact is not strong enough the induced noise will not produce benefits in the predictions. We also note that the simulation study was constructed in a way that satisfies model assumptions, in particular conditional sample exchangeability. We highlight that in practical applications, these assumptions cannot be verified \citep{yang2024doubly}. Overall, these results highlight the importance of considering the effect of covariate shift, as ignoring it may lead to worsened predictions when such shift is pronounced. However, depending on the problem of interest, such explorations may reveal minimal (if any) improvement, providing supporting evidence for a simpler, unweighted model.

\section{Application} \label{sec:data}

To illustrate the proposed methodologies for prediction and transportability, we apply them to the 2015 data of the OECD's Program for International Student Assessment (PISA) for Belgium. PISA is a worldwide study measuring 15-year-old students' competencies in reading, mathematics, and science, as well as selected additional domains. These assessments are intended to provide internationally comparable data on education systems that can help improve education policies and outcomes.

\subsection{Data}

The structure of the Belgian PISA 2015 data offers an interesting and relevant example of forecasting missing scores on a standardised student assessment. All regions of the country participated in the general assessment of PISA, while only one region participated in the optional financial literacy assessment. In the Belgian federal state, the Communities are responsible for education\footnote{Fore more information on financial education in Belgium we refer to \cite{DeWitteeducationFlanders}.}: The Flemish Community (\textit{Flanders}), the French Community and German Community (\textit{Wallonia}).\footnote{To simplify, we use Flanders to denote the Flemish Community (responsible for Flemish-speaking schools in Flanders and Brussels), and Wallonia comprising both the French Community (responsible for French-speaking schools in Wallonia and Brussels) and the German Community (responsible for the minority German-speaking schools in Wallonia).} While the first participated in both the general and financial literacy assessment, the latter two participated only in the general assessment. This structure of the data allows us to use a common set of predictors for all regions from the general assessment in order to predict financial literacy outcomes for students in the region that did not participate in this part of PISA.\footnote{In the eight OECD countries participating in the PISA financial literacy assessment 2015, Belgium and Canada are the ones with the most remarkable differences in regional assessments.}

We choose the Belgian case to illustrate the forecasting methodology, since the regions of the country have many similarities and common policies that justify the assumption of similar underlying relationships between the predictors and the outcome variable. At the same time, the Belgian regions differ in population characteristics, which suggests that the missing scores in Wallonia are unlikely to be identical to those in Flanders. In the next two Subsections, we will describe the set of predictors (Subsection \ref{subsec:predictors}) and the outcome (Subsection \ref{subsec:outcome}) in more detail.

\subsection{Outcome Variable}\label{subsec:outcome}

The financial literacy score in PISA 2015 is based on a test of financial knowledge with 43 items in four content categories: (i) money and transactions; (ii) risk and reward; (iii) planning and managing finances; and (iv) the financial landscape \citep{OECD2017a}.\footnote{The FLS in PISA 2015 are reported as a set of ten plausible values. For each participating country or economy, 33\% of the students who completed the general PISA assessment were tested on financial literacy \citep{OECD2017a}. The plausible values were then constructed for all students participating in the general PISA based on item response theory and latent regression. In the case of Belgium, the plausible values are available for all participating students in the Flemish region, but are missing for students from Wallonia. In the following, we use the plausible value $PV1FLIT$ for the analysis, since any bias caused by using a single plausible value instead of all ten simultaneously is arguably negligible \citep{Gramatki2017}. } This data structure creates a natural test case. Since we observe math and reading scores in both regions, we can validate our method before applying it to the missing financial literacy scores.

In the case of Belgium, there are 9,651 observations on the general assessment from all Belgian regions, but the FLS are only available for the 5,675 students from Flanders, while the FLS of the 3,976 students from Wallonia are missing. Flanders thus represents our ``source'' population in which we observe the FLS $Y^{obs}$. Wallonia represents the ``target'' population for which we predict the missing FLS, $Y^{mis}$, based on the common set of predictors from the general assessment $\bX$. Table \ref{tableoutcome} in Appendix \ref{subsec:app-data} provides summary statistics of the outcome variable. On average, Flemish students score 541 points, which corresponds to the PISA proficiency level 3 out of 5 levels.\footnote{These statistics may differ from official PISA reports, which rely on the full set of 10 plausible values, sampling weights, and replicate weights.} Figure \ref{fig:outcome} in Appendix \ref{subsec:app-data} shows the distribution of FLS in the Flemish data. 12\% of Flemish students fail to reach the baseline level 2 of 400 points or more which the OECD defines as the level necessary to participate in society \citep{OECD2017a}. Figure \ref{fig:PISAlevels} in Appendix \ref{subsec:app-data} provides an overview of the required knowledge corresponding to the five PISA proficiency levels.

\subsection{Predictors}\label{subsec:predictors}

In contrast to regression analyses, in which multicollinearity of regressors needs to be avoided, a large number of (potentially correlated) predictors can be used for forecasting in machine learning \citep{vaughan2005using, makridakis2008forecasting, shmueli2010explain}. We therefore select a broad set of predictors from the general assessment of PISA 2015. Table \ref{tab:variables} in Appendix \ref{subsec:app-data} provides an overview of the variables used as predictors. The set of predictors includes: (i) students' background characteristics; (ii) proxies of the socioeconomic status of students; (iii) indicators of student achievement and attitudes; (iv) school characteristics.

To account for students' background, we include gender, grade, age, language and study track \citep{Cannistra2024}. Existing studies commonly find FLS to be highly correlated in particular with math and reading performance \citep[e.g.,][]{Mancebon2019,Riitsalu2016}. In the case of Flanders, the correlation of FLS with math and reading scores amounts to 0.8, which is slightly higher than the OECD average of 0.75 \citep{UniversiteitGent2017}. As such, we include the PISA math score, as well as the reading score and additional variables related to student achievement, such as grade repetition, study time, and instruction time. Given that personality traits and attitudes have been found to matter for financial literacy as well \citep{Longobardi2018, Pesando2018}, we also include students' anxiety and motivation levels recorded using standard tests in the PISA survey. 

Another major factor associated with FLS is the family background, such as parental characteristics, language spoken at home, or immigration background \citep[e.g.,][]{Mancebon2019,Gramatki2017, compen2021impact}. To approximate the socioeconomic status of a student, we use the indicators of educational and economic possessions of the family, the number of books in the home, the immigration background of the student, the mother's and father's education and job, as well as a variable capturing perceived parental emotional support \citep[e.g.,][]{maldonado2022effects, compen2023improving}.

Finally, studies of PISA financial literacy outcomes commonly include school-level variables \citep[e.g.,][]{Pesando2018,Cordero2018}. This is particularly relevant in a context of high segregation at the school level, such as in Belgium \citep[e.g.,][]{debeckker2021educationflanders,maldonado2021impact}. We include a number of school characteristics from the questionnaire for school principals, such as the school's location, size, and autonomy. As indicators of teaching quality, we use the student-teacher ratio, class size, and teacher professional development \citep[e.g.,][]{compen2021impact}. We also use school-level indicators of socioeconomic status, such as the share of students with a different home language, special needs, or a socioeconomically disadvantaged home, the number of available computers, and the shortage of educational material.

Figure \ref{fig:missingness} in Appendix \ref{subsec:app-data} shows the missing observations in the variables used in the analysis. No clear patterns of missingness appear across the predictors. We can therefore assume the observations to be Missing-Completely-at-Random \citep{little2019statistical} and we proceed with multiple imputations using the Fully Conditional Specification (FCS) developed by \cite{buuren2010mice} and implemented in the \texttt{R} package \texttt{MICE}.

\begin{figure}[h]
    \centering
    \begin{subfigure}[b]{0.48\textwidth}
    \includegraphics[width=\textwidth]{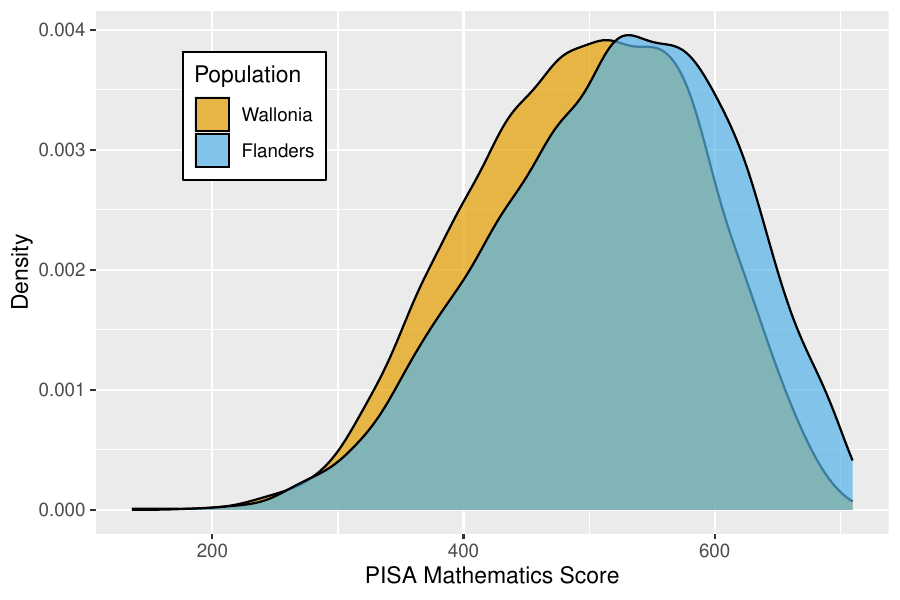}
    \end{subfigure}
    \begin{subfigure}[b]{0.48\textwidth}
    \includegraphics[width=\textwidth]{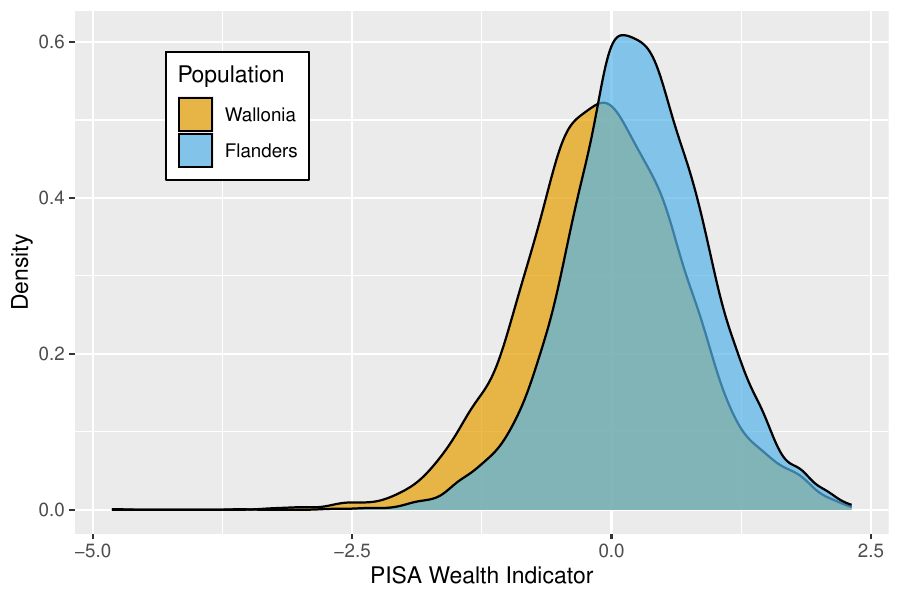}
    \end{subfigure}
    \caption{{\footnotesize(Left) Distribution of PISA math  Scores in Flanders (blue) and Wallonia (yellow).\\ (Right) Distribution of PISA Wealth Index in Flanders (blue) and Wallonia (yellow).}}
    \label{fig:mathscore}
\end{figure}

As shown in Table \ref{tablesummarystats} in Appendix \ref{subsec:app-data}, a series of T-tests reveals significant differences in means on background characteristics of students in Flanders and Wallonia. The Belgian sample is balanced across regions in terms of gender, age, and parents' characteristics. However, potentially important variables for financial literacy, such as math and reading scores, as well as socioeconomic status, are significantly different in the two regional samples. Figure \ref{fig:mathscore} shows that the distribution of math scores is shifted to the right for Flanders compared to Wallonia. A similar pattern is observed regarding the socioeconomic status of students in the two samples, as approximated by the PISA wealth indicator, which summarizes the economic possessions of the family. Figure \ref{fig:mathscore} shows that, similarly, the distribution of wealth is slightly shifted to the right in Flanders compared to Wallonia. 

The presence of such shifts in the covariate distributions motivates the need for potential adjustment to ensure transportability of predictions to the target population, namely, Wallonia. As the FLS are unobserved, the prediction error cannot be obtained for the outcome of interest. Given the high correlation between the outcome and math and reading scores, and given the standardized way all these outcomes are recorded, these can be considered to be proxies of FLS. We therefore run preliminary analyses on the math scores and reading scores, observed for students both in Flanders and Wallonia. 

\subsection{Preliminary Analyses on Proxy Outcomes}

We compare predictions for math scores and reading scores obtained under the standard BART algorithm to those arising from a re-weighted BART model. The overlap scores, used to obtain the individual weights, are estimated using a logistic regression model trained on a subset of pre-selected covariates that were found to be predictive of the outcome. The choice was made to avoid overfitting and the diagnosis of artificial separation based on predictors having little to no influence on the score values. More details can be found in Appendix \ref{sec:covariateshiftappendix}. 

To assess the performance of the two predictive models, we use 10-fold cross-validation to obtain the root-mean-squared-error (RMSE), mean-absolute-error (MAE) and coefficient of determination ($R^2$) on the left-out samples. In order to mimic the FLS case, where outcomes are unobserved for the Walloon population, both the unweighted and weighted BART models are trained exclusively using the Flanders population. The target is also split into folds to ensure that predictions are made on data that was not seen by the model, even if only at the weight estimation level. We also include results for a cross-validated model trained using Walloon data. This serves as a ``best-case scenario'' reference that yields the estimates that one could get if the outcomes were not missing. The average of the performance metrics across the 10 folds is reported in Table \ref{table:MATHREAD_target}.  

\begin{table}[H]
\centering
\small
\caption{Summary of the performance of BART for math and reading scores in Wallonia.}
\label{table:MATHREAD_target}
\begin{tabular}{lccc}
\toprule
\textbf{Model Setting} & \textit{RMSE} & \textit{MAE} &  $R^2$ \\
\midrule
\textit{Math Score Outcome} & & & \\
\quad Unweighted     &    51.642    &  41.112      &   0.674    \\
\quad Weighted       &    52.424    &   41.797     &   0.664    \\
\quad Target-trained &      43.437  &    34.658    &    0.769   \\
\addlinespace
\textit{Reading Score Outcome} & & & \\
\quad Unweighted     & 52.131       &   41.404     &    0.872   \\
\quad Weighted       &     57.866  &     46.083   &   0.780    \\
\quad Target-trained &    47.550    &  37.746      &   0.906    \\
\bottomrule
\end{tabular}
\end{table}

The analysis of the two proxies reveals that covariate shift adaptation does not improve the model performance on the unobserved target data. Indeed, both for math and reading scores, the errors are slightly lower under the unweighted BART model, which directly applies predictions to the Walloon data without concerns about transportability. 

We identify several factors that are potentially contributing to this behavior. The most important concern is that the distribution of overlap scores in the Flemish population is relatively concentrated near 1, see Figure \ref{fig:overlap_dists} in Appendix \ref{sec:covariateshiftappendix}. As a result, a large number of weights carry small values, effectively removing a large number of observations from the training sample, some of which are certainly informative. 

The reweighting scheme can be thought of as bias correction at the expense of variance inflation. Therefore, under extreme weights, high variance is introduced for a small improvement in bias. Alternative techniques for the estimation of overlap scores were also considered to improve performance. However, no type of weighted model was found to surpass the performance of the unweighted one.

While balancing weights does not improve predictions with the covariate shift issue in this application, this analysis illustrates reasonable robustness of the standard BART model for prediction. Indeed, while the errors are lower for the model trained using Wallonia data (in particular regarding the math score outcome), BART trained on Flanders data still performs reasonably well. Furthermore, the concentration of the overlap scores towards larger values is not necessarily revealing of fundamental overlap issues, but rather an artifact of certain predictors, or combinations thereof, that lead to good discrimination between populations. A two-dimensional representation of the covariate space can be obtained using UMAP \citep{mcinnes2018umap}, as illustrated in Appendix Figure \ref{fig:umap}. The lack of identifiable clusters by population, with comparable scatter from Wallonia and Flanders observations across the representation space, tends to support the assumption of reasonable overlap. The BART model, therefore, should not overly rely on extrapolation, and a meaningful analysis of the FLS can still be conducted.

We also conducted robustness checks evaluating alternative weighting strategies, including weight trimming (winsorization) and PCA-based dimension reduction. These analyses, presented in Appendix~\ref{appendix:weighting}, confirm that even with these adjustments, the unweighted BART model outperforms all weighted variants in our setting.


\section{Applied Results} \label{sec:results}

Our analysis proceeds in three stages: first, establishing BART's prediction accuracy, then testing whether covariate shift adjustment improves predictions, and, finally, identifying student characteristics associated with low financial literacy (vulnerable students).

\subsection{Predictive Performance of BART and Random Forest}

Many studies have shown that BART performs well in various predictive tasks \citep{murray2021log,linero2018bayesian1, linero2018bayesian2, hernandez2018bayesian, bargagli2024machine, bargagli2021supervised}. In our application, we assess the performance of BART through 10-fold cross-validation and compare it with random forest (RF) \citep{breiman2001random}, one of the most widely used algorithms for prediction in social sciences \citep{athey2018impact, bargagli2021supervised}. Table~\ref{table:performance} presents the comparative results. BART achieves slightly lower RMSE and MAE, and marginally higher $R^2$ compared to RF. However, 
we acknowledge that these differences are modest, and both methods demonstrate comparable predictive performance in this setting. 

In addition to measures of predictive performance, we also include measures of uncertainty, evaluated on the training samples and averaged across 10 folds. For the BART model, we obtain 95\% posterior predictive intervals, based on MCMC samples from the posterior predictive distribution. We calculate the proportion of observations for which the FLIT scores falls in the interval, which we report as coverage. We also report the average interval width across observations. The Random Forest algorithm is not as easily amenable to uncertainty quantification, and many approaches tend to focus on conditional mean intervals rather than prediction intervals, failing to account for random error. To obtain prediction intervals comparable to those obtained from BART, we follow the approach of \cite{Zhang2019} which is based on the empirical distribution of out-of-bag prediction errors. We then report coverage and average interval width at the 95\% level as well. The BART average interval width is smaller than the one obtained for RF, but the approach suffer from slight undercoverage (0.941). However, we find the flexibility and convenience of BART as a tool for assessing uncertainty surrounding any quantity of predictive or inferential interest------features 
central, in particular, to our outlier detection analysis in Section~\ref{sec:results}---to outweigh these marginal differences.

\begin{center}
\begin{table}[H]\centering
\small
\caption{Summary Statistics of the Performance of the ML techniques}
\label{table:performance}
\begin{threeparttable}
\begin{tabular}{l*{1}{ccccc}}
\toprule
                &     \textit{RMSE} & \textit{MAE} &  $R^2$ & Coverage & Avg. Interval Width\\
\midrule
BART &    57.968 &    45.610 &    0.733 & 0.941 & 224.206\\
Random Forest &   58.556 &   45.947 &     0.727 & 0.951 & 232.607\\
\bottomrule
\end{tabular}
\end{threeparttable}
\end{table}
\end{center}

\begin{figure}[h]
    \centering
    \includegraphics[width=0.85\textwidth]{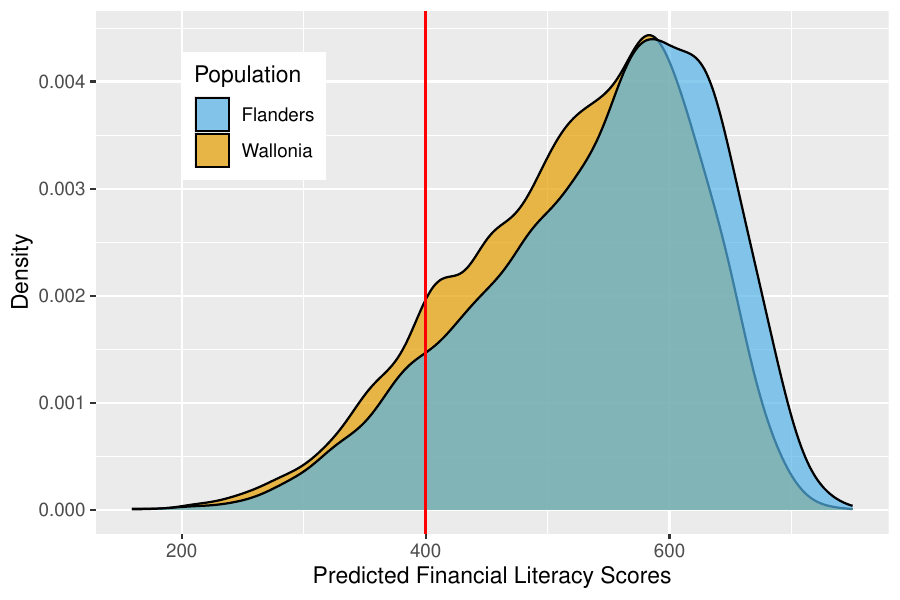}
    \caption{Predicted FLS for Flanders (light blue) and Wallonia (orange). The red line indicates the threshold of the baseline level of proficiency in financial literacy. OECD suggests that students above this threshold of 400 points have financial literacy levels that are sufficient to participate in society \citep{OECD2017a}.}
    \label{fig:predictions}
\end{figure}

Figure \ref{fig:predictions} depicts the posterior predicted values for FLS for students in Flanders (light blue) and Wallonia (orange), while Table \ref{table:fls} reports summary statistics for the same predictions. From Table \ref{table:fls} we observe that the mean posterior predictive FLS for students in Flanders is higher  than for those in Wallonia, 541.4 compared to 516.6. Considering the minimum and maximum values for Flanders and Wallonia, we note that the scores of the students in Flanders are more centered around the mean than those in Wallonia. The same conclusion can be drawn from Figure \ref{fig:predictions} where the peak of the light blue area (Flanders) is to the right of the peak of the orange area (Wallonia).

The OECD provides a way to interpret the scores. In particular, based on OECD computations, a difference of 40 points equals an entire school year of learning. Hence, based on this, we can deduce that, on average, students in Flanders are roughly half a school year ahead of students in Wallonia according to their standardized knowledge in financial literacy.

\begin{center}
\begin{table}[H]\centering
\small
\caption{Summary Statistics of the Predicted FLS}
\label{table:fls}
\begin{threeparttable}
\begin{tabular}{l*{1}{cccccc}}
\toprule
                &     Mean&       SD&  Minimum&   Median&  Maximum&      N\\
\midrule
Predicted FLS Flanders &   541.4 &   96.8 &    171.2 &   560.1 &   750.1 &     5675\\
Predicted FLS Wallonia &   516.7  &   93.9 &   159.1 &   528.4 &   730.7 &     3976\\
\bottomrule
\end{tabular}
\end{threeparttable}
\end{table}
\end{center}

As earlier established, the quality of the predictions of FLS in Wallonia is not directly testable as the outcome is unobserved in this region. The analysis conducted in Section \ref{sec:data} would hint that the quality of the predictions is a bit lower than that of those obtained on cross-validated samples from Flanders, but still provides meaningful insights into the underlying truth. 

\subsection{Results Under Domain Adaptation}

Since the financial literacy scores are unobserved in the Wallonia region, we adopt the covariate shift methodology introduced in Section \ref{sec:covariateshift}. The balancing weights are learned by estimating the overlap scores using a logistic regression model trained using a selected subset of the available predictors. To avoid an overly discriminating model for overlap, we only include covariates that carry high explanatory power for the prediction of FLS. The specific set is determined by considering variable importance. This approach ensures that the potential shift in distributions from source to target is accounted for in the covariates that have the strongest influence on predictions, while avoiding artificial separation by covariates that are uncorrelated with the outcome.

BART was earlier identified as the preferred predictive model for the FLS prediction task. The balancing weights for the Flanders data are learned using the full dataset and are then incorporated by scaling the individual observation variances in the BART likelihood to adjust for covariate shift. The missing FLS for students in Wallonia are predicted using this modified model. Specifically, under the Gaussian model $y_i \mid f, \sigma^2 \sim \mathcal{N}(f(\mathbf{x}_i), \sigma^2)$, weighting observation $i$ by $\omega_i$ is equivalent to assuming heteroskedastic variance $\sigma^2_i = \sigma^2 / \omega_i$.

\begin{figure}
    \centering
    \includegraphics[width=0.85\textwidth]{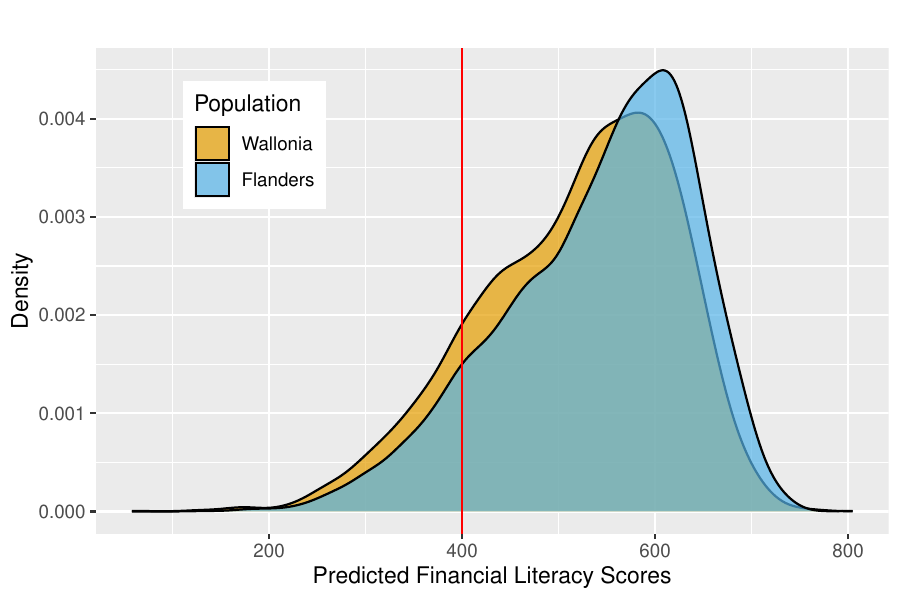}
    \caption{Predicted FLS for Flanders (light blue) and Wallonia (orange) under the weighted BART model. The red line indicates the threshold of the baseline level of proficiency in financial literacy.}
    \label{fig:predictions_flit_weighted}
\end{figure}

\begin{center}
\begin{table}[H]\centering
\small
\caption{Summary Statistics of the Predicted FLS under Domain Adaptation}
\label{table:fls2}
\begin{threeparttable}
\begin{tabular}{l*{1}{cccccc}}
\toprule
                &     Mean&       SD&  Minimum&   Median&  Maximum&      N\\
\midrule
Predicted FLS Flanders &  540.9  &   98.9 &   136.4  &  559.0  &   804.6 &  5675   \\
Predicted FLS Wallonia &    517.9 &   101.0 &  57.8  &  534.2  &   771.3 &   3976  \\
\bottomrule
\end{tabular}
\end{threeparttable}
\end{table}
\end{center}

Figure \ref{fig:predictions_flit_weighted} represents the distribution of the posterior predicted values for FLS under the covariate shift framework, and Table \ref{table:fls2} represents the summary statistics for these predictions. These can be compared to their counterparts under the standard BART model, given in Figure \ref{fig:predictions} and Table \ref{table:fls}. The mean and median are very similar for both Flanders and Wallonia students, but the spread in predicted values increases, in particular for the unobserved group. This difference is notable in the standard deviation as well as the minimum and maximum values. Furthermore, inspection of the density seems to indicate a larger discrepancy in score distribution between regions. Under the BART model, the densities display a location shift but overall similar shapes. On the other hand, the results obtained under the reweighting scheme display a similar variability in central tendency, but additionally, the score distribution for Wallonia has heavier tails than that for Flanders.

\begin{figure}
    \centering
    \includegraphics[width=0.7\textwidth]{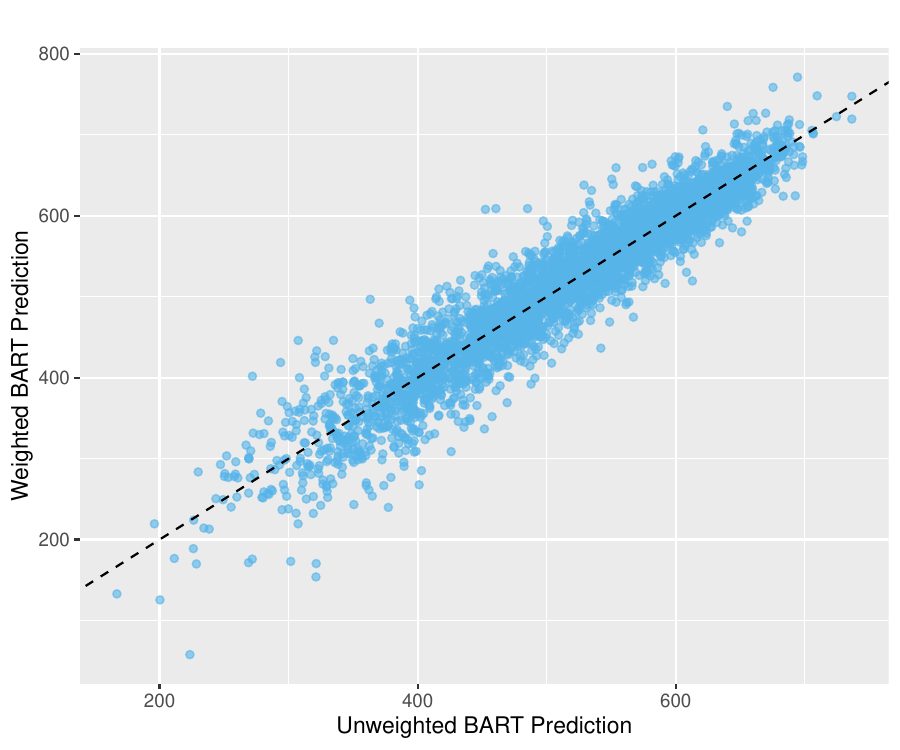}
    \caption{Comparison of predicted FLS scores under the weighted and unweighted BART models. The dotted line corresponds to the identity: observations are centered around it, indicating minimal discrepancy in predictions between the two models.}
    \label{fig:flit_weighted_v_unweighted}
\end{figure}

Figure \ref{fig:flit_weighted_v_unweighted} displays a comparison of individual predictions under the two modeling frameworks. One would expect the points to generally follow the straight line $y = x$, which appears to overall be the case. Predictions from the weighted BART model appear to be greater than those from the unweighted one at high FLS values, and vice versa for low values. This is consistent with the previously observed difference in the spread of scores. 

Ideally, covariate adjustment would improve the performance of predictions for the target study, that is, the student scores in the Wallonia region. In Section \ref{sec:data}, it was, however, diagnosed that, for this dataset,  standard BART tends to overperform the modified version. Our weighted BART approach is motivated by the semiparametric efficiency theory for covariate shift adaptation. However, recent work by \cite{kennedy2023towards} on optimal doubly robust estimation reveals important caveats. When the transported outcome function is sufficiently smooth, simpler unweighted approaches can achieve oracle efficiency. Therefore, subsequent analyses will hence be conducted using the former model. 

\subsection{Results for Vulnerable Students Detection}

Policymakers, facing budget constraints, often want to target those who are most in need of an intervention. In cases where the values of a certain outcome of interest are observed, population groups at risk can be identified by considering the outcome's distribution. However, often such outcomes may be hidden or unobserved for a part of the source population. In such scenarios, machine learning predictions may furnish a valuable tool to policymakers. In the case of our application, we face a prediction problem: imagine that policymakers want to provide additional learning material to more vulnerable students in a region where FLS are not observed. Machine learning can provide a useful tool to draw such predictions.

In order to detect the most vulnerable students (i.e., those with low predicted FLS), we use the procedure introduced in Section \ref{subsec:bart}. Uncovering subpopulations of students that differ in terms of the distribution of a certain outcome is central to boosting schools' and teachers' effectiveness, and various strategies have been applied to this task \citep{masci2019semiparametric}. Bayesian machine learning has been applied to discover the relationship between financial literacy and self-reported economic outcomes \citep{puelz2022financial}. However, to our knowledge, this is the first time that a supervised machine learning technique is used to predict students' FLS.

In a previous study on OECD data, \cite{DeBeckker2019} identify four groups of adults with different financial literacy levels through an unsupervised machine learning algorithm (i.e., k-means). The advantage of the use of machine learning is clear, as it allows for partitioning a large heterogeneous group of people into subgroups according to their FLS. However, while simple unsupervised machine learning algorithms such as k-means are useful to get a first impression of the different subgroups and their socioeconomic characteristics, they do not provide immediate information on the robustness of the outcomes. Moreover, these simple algorithms are not capable of making predictions for out-of-sample regions. This is relevant when dealing with standardized tests, such as PISA, which may only cover subpopulations of a country, such as specific regions.

First, we train BART on the sample of Flemish students for which we have data on FLS. Next, we predict the FLS for both the sample of students in Flanders and Wallonia. After that, we compare the posterior predicted values for each student with the mean of the predicted posterior values. Finally, we detect the students for whom these values lie outside the credibility intervals for the mean values. Here, we define outliers as those observations with predicted FLS values smaller than two standard deviations from the mean, following \cite{miller1991reaction}. In Appendix \ref{appendix:outliers}, we check the results we would obtain when defining the outliers as observations with predicted FLS two absolute deviations smaller than the median as suggested by \cite{leys2013detecting}.\footnote{Such results are not significantly different from the ones that are depicted in this Section. Thus, we argue that our results are robust to different definitions of outlying predictions.}

To get a better understanding of which variables best explain low predicted FLS, we generate a dummy variable that assumes the value of 1 if the student has a low predicted FLS and the value of 0 otherwise. Finally, we built a series of conditional inference trees \citep{hothorn2006unbiased} to have a descriptive sense of which are the drivers of low financial literacy.  This fit-the-fit procedure has been widely applied in recent years to obtain interpretable results for the main drivers in the heterogeneity of an outcome \citep{lee2022discovering, bargagli2022heterogeneous, bargagli2020cre}.

\begin{figure}[H]
    \centering
    \includegraphics[width=1\textwidth]{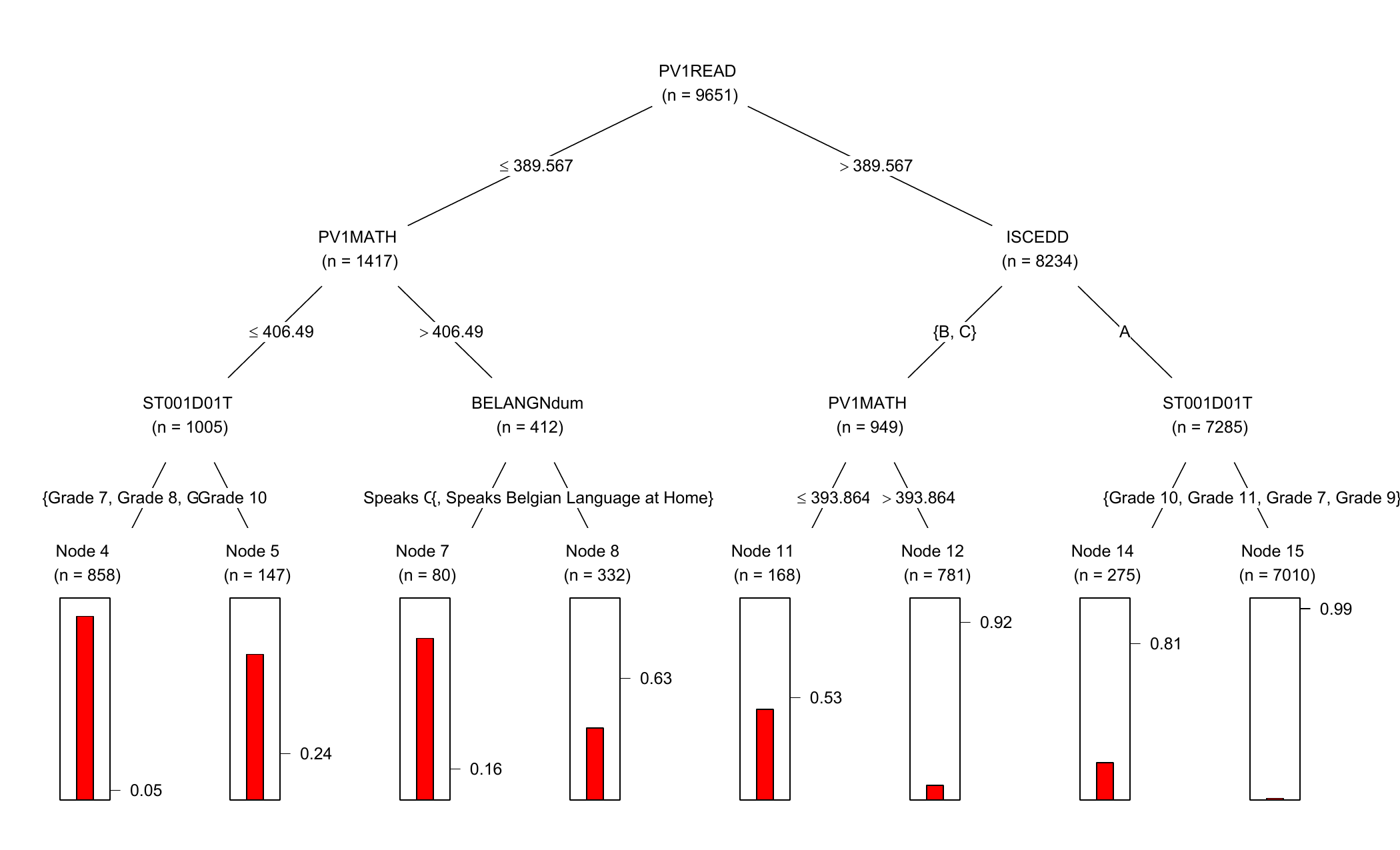}
    \caption{Conditional tree for the entire sample. Within each leaf is depicted in red the histogram of the percentage of units that have a low financial literacy score, and next to it the percentage of units with not-low FLS within the same leaf.}
    \label{fig:ctree_general}
\end{figure}

Figure \ref{fig:ctree_general} depicts the conditional tree for the overall sample, Figure \ref{fig:flanders_vs_wallonia1} shows the conditional trees for Flanders and Wallonia, respectively. In the Appendix, we report Figure \ref{fig:ctree_grades1}, which depicts the trees for students in grades 7 to 9 and 10 to 12, and Figure \ref{fig:ctree_vocational1}, which shows the results for the trees for students in general and vocational education.

As shown in Figure \ref{fig:ctree_general}, reading (\texttt{PV1READ}) and math  (\texttt{PV1MATH}) scores, study track (\texttt{ISCEDD}), grade (\texttt{ST001D01T}) and the language spoken at home (\texttt{BELANGN}) are important variables distinguishing groups of students in our entire sample. For students from the lower grades (grade 7 to 9) with lower scores on reading ($\leq389.567$) and math ($\leq406.49$) the predicted probability of low FLS is 95 percent. A higher PISA math score ($>406.49$), but not being a native speaker, results in a predicted probability of 84 percent of having low FLS. On the opposite side, students with a reading score above baseline ($>389.567$), following general programs designed to give access to tertiary education, have---depending on their grade level---a predicted probability of respectively only 19 or 1 percent of having low FLS.  

In Figure \ref{fig:flanders_vs_wallonia1}, we split by region. In the tree in the left panel of the Figure, we note that in Flanders, the largest predictive proportion of students with low FLS are again situated among those with lower scores for reading ($\leq386.732$) and math ($\leq405.291$). In the group with a low reading score ($\leq386.732$) and a higher math score ($>405.291$), immigration status seems to matter (\texttt{IMMIG}). Among those who are first-generation migrants, the predictive probability of having low FLS is 82 percent. In Wallonia, the proportion of students with low FLS is again the largest (94 percent) among those with low reading ($\leq385.283$) and math  scores ($\leq406.49$). In Wallonia (tree in the right panel), mother's education (\texttt{MISCED}) seems to play a distinctive role. Students with better reading and math  scores and with a highly educated mother only have a predictive probability of 1 percent of low FLS.

\begin{figure}[H]
    \centering
    \begin{subfigure}[b]{0.48\textwidth}
    \includegraphics[width=\textwidth]{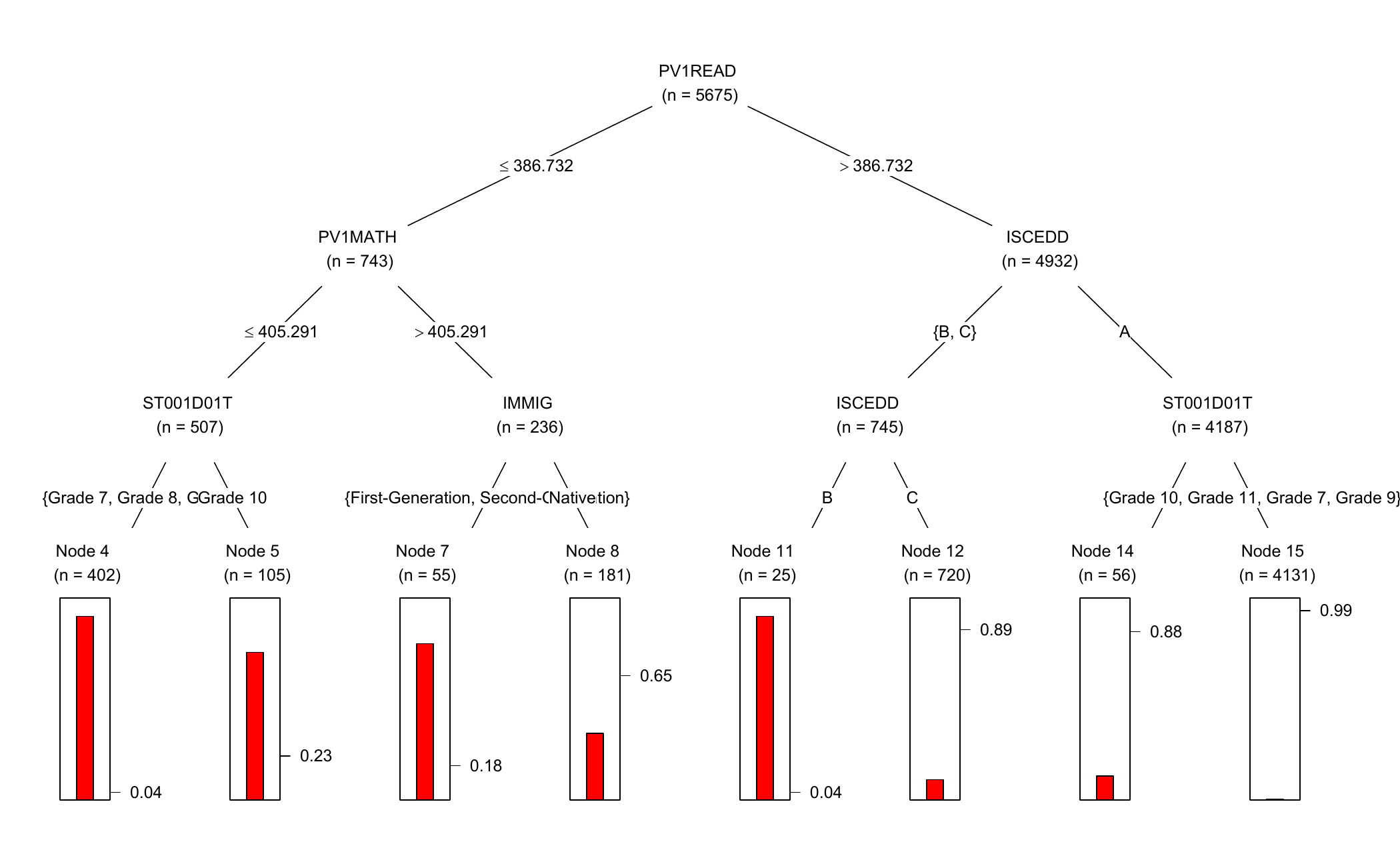}
    \end{subfigure}
    \begin{subfigure}[b]{0.48\textwidth}
    \includegraphics[width=\textwidth]{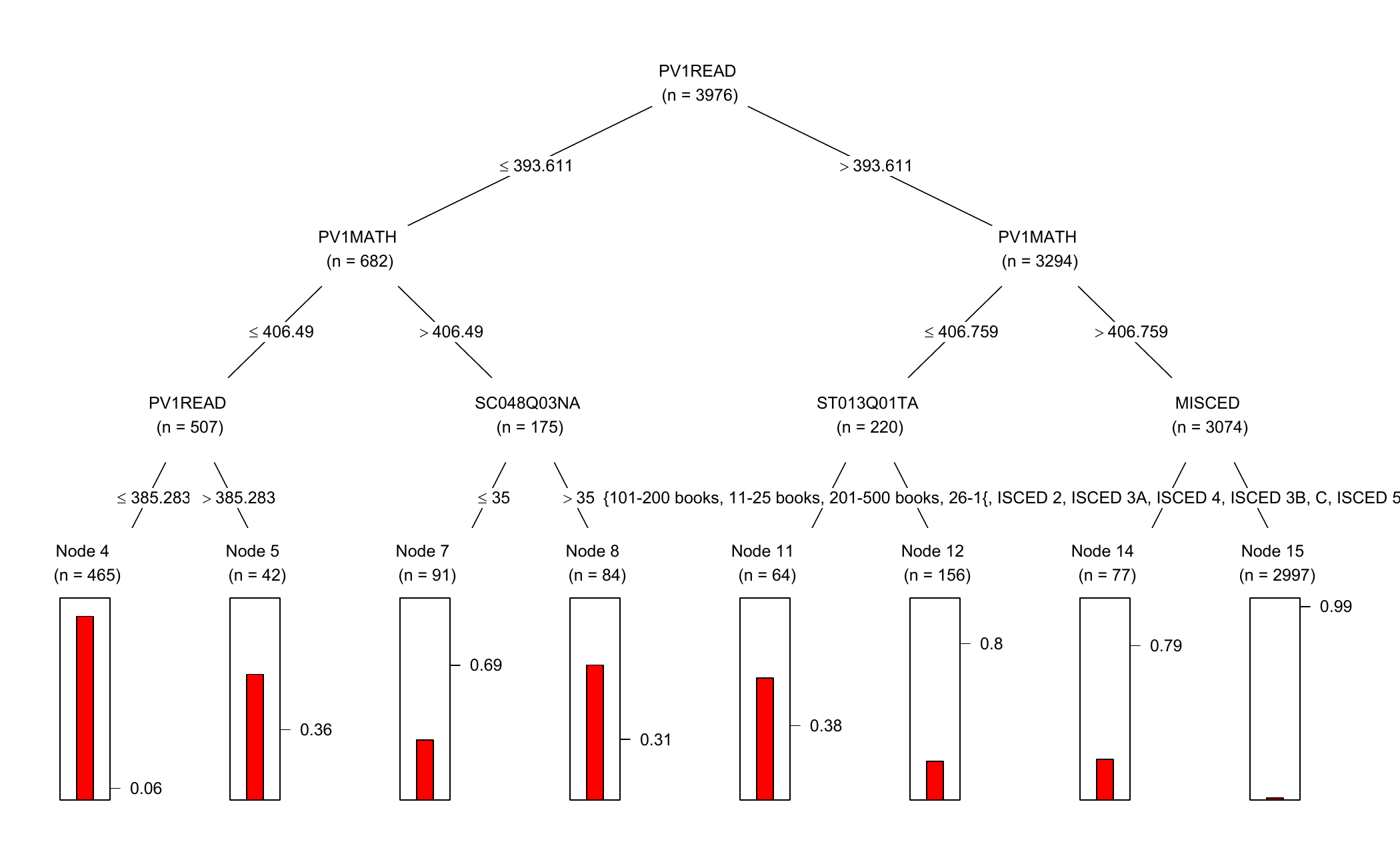}
    \end{subfigure}
    \caption{\footnotesize(Left) conditional tree for Flanders. (Right) The corresponding tree for Wallonia.}
    \label{fig:flanders_vs_wallonia1}
\end{figure}

\section{Discussion and Conclusions} \label{sec:discussion}

In this paper, we introduce a novel double machine learning technique for the prediction of student financial literacy scores. In the PISA dataset, FLS was measured in the Flanders region but unobserved in Wallonia. A standard but naive approach would train a prediction model using the available source data and use it to directly obtain estimates for the unobserved quantity of interest. However, this fails to account for potential discrepancies in population characteristics between the source and target, and may introduce bias. Therefore, we propose a two-step approach in which we first obtain overlap scores for each observation used to train the predictive model, corresponding to their respective probabilities to belong to the source population. Overlap weights are then constructed to emphasize the contribution of points whose covariates resemble the target more closely. 

This proposed approach effectively builds a modified training set that is aligned, through importance weighting, with the population for which predictions are being estimated. This contribution is critical in the case of covariate shift, as resulting estimates may otherwise be inaccurate, leading to flawed conclusions to inform policy making. For our approach, we adopt a Bayesian machine learning model, BART, which depicts a strong predictive performance with and without the incorporation of overlap weights.  A fully Bayesian approach would incorporate the learning of the latent weights in a probabilistic framework. We leave this to future research.

The usage of a Bayesian machine learning method, BART, serves three purposes in our framework. First, it provides principled uncertainty quantification through the posterior predictive distribution, enabling credible intervals for transported predictions. Second, it naturally accommodates observation weighting through likelihood reweighting---as explained in Section 2.5, weighting observation $i$ by $\omega_i$ under a Gaussian likelihood is equivalent to scaling its variance as $\sigma^2_i = \sigma^2 / \omega_i$. Third, it enables the outlier detection analysis in Section 2.6, where draws from the posterior predictive distribution are used to identify observations with extreme predicted values. Our current implementation treats the estimated overlap scores as fixed in the second stage. Again, a fully Bayesian extension would jointly model population membership and the outcome: this integration represents a promising direction for future methodological development.

Our simulation study demonstrates that importance weighting improves predictive performance when covariate shift is pronounced. However, when shift is small or moderate, the variance introduced by weighting can outweigh its benefits. These findings highlight a practical trade-off: ignoring covariate shift risks degraded predictions when shift is severe, yet correcting for it when shift is minimal may introduce unnecessary noise. In the latter case, the principle of parsimony favors simpler, unweighted models \citep{bargagli2020causal}. Our empirical application illustrates this scenario.

In our application, we start by applying standard machine learning techniques to predict FLS for students in the Wallonia region of Belgium, where FLS are unobserved. Next, we adjust our model through the balancing weights to implement domain adaptation, dampening the effect of potential covariate shift. Finally, we identify the characteristics of students with predicted FLS lower than the mean. The first stage of our analysis provides evidence that the BART approach outperforms the traditional RF approach on a number of selected performance measures (i.e., smaller RMSE and MAE, higher $R^2$) for FLS predictions. We find that the predicted values for FLS for students in Flanders are, on average, somewhat larger than those for students in Wallonia. Nevertheless, the distribution of FLS in Flanders is more extreme (i.e., lower minimum, larger maximum) compared to Wallonia. 
 
Our analysis using the reweighted BART model does not display improvement relative to the baseline framework. An analysis using math and reading scores as proxy variables reveals little effect of covariate shift. Instead, domain adaptation introduces additional noise, limiting the improvement in the models' predictive performance. The presented results reveal that BART arises as a robust model for transporting predictions from the source to the target set, validating the reasonable accuracy of the obtained predicted FLS.

Next, we estimate the posterior predicted observations available for each student, which also allows us to identify outliers (i.e., students for which the FLS are situated outside the 95\% credibility intervals for the mean values). Our results show that the predictive probability of having low FLS is the largest for students with lower scores on reading and math in the PISA test. This corroborates the findings of \cite{Mancebon2019} who argue that the development of financial abilities of students is mediated by their mathematical skills. 

Another interesting observation is that the family background also has a key influence. Students with the largest predicted low FLS are often students from families where the school language is not spoken at home. Particularly in Flanders, being a first-generation immigrant increases the probability of having low FLS. The educational background of parents is another important predictor. These observations are in line with existing literature suggesting that students at risk of low FLS are often situated among those with a lower socioeconomic background \citep{Gramatki2017, Riitsalu2016,DeBeckker2019}. 

From a policy perspective, our novel approach is interesting as it not only allows us to predict outcomes for certain countries (or regions within countries, as in the case of our application) with missing data, but also introduces the possibility of detecting observations that fall outside the mean outcomes. This enables policymakers to identify which factors drive low results. We applied our model to PISA financial literacy data; however, the same approach can also be applied to other large administrative datasets. 

\section*{Acknowledgments}

Falco J. Bargagli-Stoffi acknowledges support from  Amazon Web Services grant on ``AI/ML  for Identifying Social Determinants of Health''. Kristof De Witte acknowledges financial support from the Horizon Europe grant BRIDGE (grant 101177154).  Views and opinions expressed are however those of the author(s) only and do not necessarily reflect those of the European Union or the granting authority. Neither the European Union nor the granting authority can be held responsible for them. 

\singlespacing
\bibliographystyle{dcu}
\bibliography{biblio}

\newpage

\appendix
\pagenumbering{arabic}
\setcounter{page}{1}
\setcounter{equation}{0}
\setcounter{table}{0}
\renewcommand{\thetable}{A\arabic{table}}
\setcounter{figure}{0}
\renewcommand{\thefigure}{A\arabic{figure}}

\begin{center}
\text{\LARGE \bf Online appendix}
\end{center}
\section{Data}\label{subsec:app-data}

\begin{table}[H]\centering
\small
\caption{Summary Statistics of the Outcome Variable}
\label{tableoutcome}
\begin{threeparttable}
\begin{tabular}{l*{1}{cccccc}}
\toprule
                &     Mean&       SD&  Minimum&   Median&  Maximum&      N\\
\midrule
Financial Literacy &   541.43&   112.16&    51.81&   555.14&   901.64&     5675\\
\bottomrule
\end{tabular}
\begin{tablenotes} 
\item \tiny{\textit{Note:} Summary statistics of Plausible Value 1 in Financial Literacy from PISA 2015 for the Flemish region in Belgium. The OECD constructs ten plausible values for financial literacy using item response theory and latent regression \citep{OECD2017a}. The analysis in this paper is based on Plausible Value 1. Consequently, the estimates may differ from official PISA reports, which rely on the full set of 10 plausible values, sampling weights, and replicate weights. SD stands for the standard deviation of the variable.}
\end{tablenotes}
\end{threeparttable}
\end{table}

\vspace{-0.5cm}

\begin{table}[H]
\centering
\caption{Variables used from the PISA data}
\label{tab:variables}
\begin{tabular}{@{}ll@{}}
\toprule
PISA Code &  Variable \\ 
\midrule
\textit{Student Characteristics} &   \\ 
\texttt{ST001D01T}         & International Grade                     \\
\texttt{ST004D01T}         & Gender           \\
\texttt{AGE}               & Age  \\
\texttt{ISCEDD}            & Study Track: ISCED Designation                   \\
\texttt{ISCEDO}            & Study Track: ISCED Orientation                    \\
\vspace{0.2cm}
\texttt{BELANGN}           & Speaks Belgian Language at Home    \\
\textit{Socioeconomic Status} &    \\  
\texttt{HEDRES}            & Educational Resources at Home    \\
\texttt{WEALTH}            & Family Wealth Index (Economic Possessions)        \\
\texttt{ST013Q01TA}        & Number of Books at Home  \\
\texttt{IMMIG}             & Immigration Status               \\
\texttt{MISCED}            & Mother's Education (ISCED)             \\
\texttt{FISCED}            & Father's Education (ISCED)           \\
\texttt{BMMJ1}             & Mother's Job (ISEI)                 \\
\texttt{BFMJ2}             & Father's Job (ISEI)          \\
\vspace{0.2cm}
\texttt{EMOSUPS}           & Parents Emotional Support      \\  
\textit{Achievement and Attitude} &     \\  
\texttt{PV1MATH}           & Plausible Value 1 in Mathematics        \\
\texttt{PV1READ}           & Plausible Value 1 in Reading           \\
\texttt{REPEAT}            & Grade Repetition  \\
\texttt{OUTHOURS}          & Out-of-School Study Time per Week  \\
\texttt{MMINS}             & Mathematics Learning Time at School           \\
\texttt{LMINS}             & Language Learning Time at School              \\
\texttt{ANXTEST}           & Personality: Test Anxiety      \\
\vspace{0.2cm}
\texttt{MOTIVAT}           & Achievement Motivation            \\   
\textit{School Characteristics} &    \\  
\texttt{SC001Q01TA}        & School Community (Location)    \\
\texttt{SC048Q01NA}   & Share of Students With a Different Heritage Language \\
\texttt{SC048Q02NA}        & Share of Students With Special Needs          \\ 
\texttt{SC048Q03NA} & Share of Socioeconomically Disadvantaged Students  \\
\texttt{SCHSIZE}           & School Size             \\
\texttt{CLSIZE}            & Class Size                           \\
\texttt{RATCMP1}           & Number of Available Computers per Student      \\
\texttt{LEADPD}            & Teacher Professional Development           \\
\texttt{SCHAUT}            & School Autonomy                 \\
\texttt{EDUSHORT}          & Shortage of Educational Material       \\
\texttt{STRATIO}           & Student-Teacher Ratio                   \\
\addlinespace  \bottomrule
\end{tabular}
\end{table}
\clearpage

\begin{center}
\begin{table}[H]
\tiny
\caption{Summary Statistics of the Predictors}
\label{tablesummarystats}
\begin{threeparttable}
\begin{tabular}{l*{3}{{c}{c}{c}}}
\toprule \addlinespace
                & Flanders&         &         & Wallonia&         &         &Difference in Means\\ \addlinespace
                &     Mean&       SD&        N&     Mean&       SD&        N&  p-value\\
\midrule
\               &         &         &         &         &         &         &         \\
\addlinespace
\textit{Student Characteristics}&         &         &         &         &         &         &         \\
\addlinespace
International Grade&     9.72&     0.52&     5675&     9.44&     0.75&     3857&    0.000\\
\addlinespace
Gender          &     1.51&     0.50&     5675&     1.51&     0.50&     3976&    0.976\\
\addlinespace
Age             &    15.85&     0.29&     5675&    15.84&     0.29&     3976&    0.725\\
\addlinespace
Study Track: ISCED Designation &     1.43&     0.81&     5675&     1.24&     0.64&     3976&    0.000\\
\addlinespace
Study Track: ISCED Orientation &     2.09&     1.00&     5675&     1.55&     0.89&     3976&    0.000\\
\addlinespace
Speaks Belgian Language at Home&     0.91&     0.29&     5595&     0.87&     0.34&     3929&    0.000\\
\addlinespace
\               &         &         &         &         &         &         &         \\
\textit{Socioeconomic Status}&         &         &         &         &         &         &         \\
\addlinespace
Educational Resources at Home&     0.27&     0.91&     5585&    -0.15&     0.90&     3930&    0.000\\
\addlinespace
Family Wealth Index (Economic Possessions) &     0.27&     0.74&     5606&    -0.04&     0.85&     3936&    0.000\\
\addlinespace
Number of Books at Home&     3.02&     1.51&     5557&     3.32&     1.53&     3905&    0.000\\
\addlinespace
Immigration Status&     1.20&     0.54&     5512&     1.32&     0.66&     3851&    0.000\\
\addlinespace
Mother's Education (ISCED)&     4.62&     1.39&     5376&     4.63&     1.53&     3774&    0.882\\
\addlinespace
Father's Education (ISCED)&     4.53&     1.44&     5221&     4.52&     1.58&     3654&    0.901\\
\addlinespace
Mother's Job (ISEI)&    47.05&    22.75&     4763&    47.30&    22.10&     3159&    0.635\\
\addlinespace
Father's Job (ISEI)&    46.21&    21.38&     4713&    46.21&    22.16&     3318&    0.996\\
\addlinespace
Parents Emotional Support&    -0.01&     0.96&     5458&     0.01&     0.96&     3801&    0.302\\
\addlinespace
\               &         &         &         &         &         &         &         \\
\textit{Achievement and Attitude}&         &         &         &         &         &         &         \\
\addlinespace
Plausible Value 1 in Mathematics&   521.61&    97.89&     5675&   494.78&    90.71&     3976&    0.000\\
\addlinespace
Plausible Value 1 in Reading&   510.95&   100.67&     5675&   490.46&    95.81&     3976&    0.000\\
\addlinespace
Grade Repetition&     0.24&     0.43&     5457&     0.42&     0.49&     3804&    0.000\\
\addlinespace
Out-of-School Study Time per Week&    14.24&    10.21&     4245&    16.27&    11.61&     3216&    0.000\\
\addlinespace
Mathematics Learning Time at School&   191.28&    86.43&     5273&   219.74&    81.86&     3668&    0.000\\
\addlinespace
Language Learning Time at School&   187.24&    84.63&     5275&   227.10&    83.49&     3665&    0.000\\
\addlinespace
Personality: Test Anxiety&    -0.30&     0.97&     5416&    -0.01&     1.01&     3775&    0.000\\
\addlinespace
Achievement Motivation&    -0.64&     0.83&     5417&    -0.28&     0.87&     3767&    0.000\\
\addlinespace
\               &         &         &         &         &         &         &         \\
\textit{School Characteristics}&         &         &         &         &         &         &         \\
\addlinespace
School Community (Location)&     2.87&     0.81&     5556&     3.24&     1.11&     3681&    0.000\\
\addlinespace
Share of Students With a Different Heritage Language&    16.41&    23.39&     5406&    26.19&    30.29&     2812&    0.000\\
\addlinespace
Share of Students With Special Needs&    19.58&    20.14&     5154&    18.97&    21.94&     3005&    0.208\\
\addlinespace
Share of Socioeconomically Disadvantaged Students &    20.02&    22.11&     5248&    33.06&    30.38&     3120&    0.000\\
\addlinespace
School Size     &   695.86&   331.42&     5405&   771.27&   327.30&     3413&    0.000\\
\addlinespace
Class Size      &    18.89&     5.14&     5592&    21.01&     3.77&     3578&    0.000\\
\addlinespace
Number of Available Computers per Student&     1.25&     0.89&     5270&     0.47&     0.33&     3251&    0.000\\
\addlinespace
Teacher Professional Development&     0.10&     0.90&     5150&     0.11&     1.07&     3433&    0.685\\
\addlinespace
School Autonomy &     0.77&     0.18&     5675&     0.59&     0.21&     3626&    0.000\\
\addlinespace
Shortage of Educational Material&     0.02&     0.87&     5488&     0.21&     0.87&     3429&    0.000\\
\addlinespace
Student-Teacher Ratio&     9.09&     3.21&     5325&     9.14&     2.66&     2847&    0.454\\
\bottomrule
\end{tabular}
\begin{tablenotes} 
\item \footnotesize{\textit{Note:} Summary statistics of all predictors used in the analysis from the PISA 2015 data for Flanders and Wallonia. SD stands for the standard deviation of the variable. The last column shows the p-value of a two-sample t-test for the equality of means across the regions of Flanders and Wallonia. Language has been grouped for Belgium (1=Dutch, 2=French, 3=German, 4=Other). These statistics may differ from official PISA reports, which rely on the full set of 10 plausible values, sampling weights, and replicate weights.}
\end{tablenotes}
\end{threeparttable}
\end{table}\label{subsec:sumstats}
\end{center}

\vspace{3cm}

 \begin{figure}[H]
    \centering
    \resizebox{1.1\textwidth}{!}{
    \includegraphics{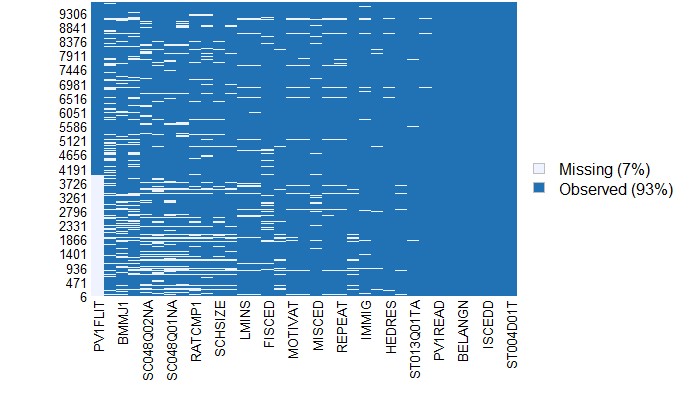}
    }
    \caption{Missingness Map}
    \label{fig:missingness}
\end{figure}

\begin{figure}[H]
    \centering
    \resizebox{0.9\textwidth}{!}{
    \includegraphics{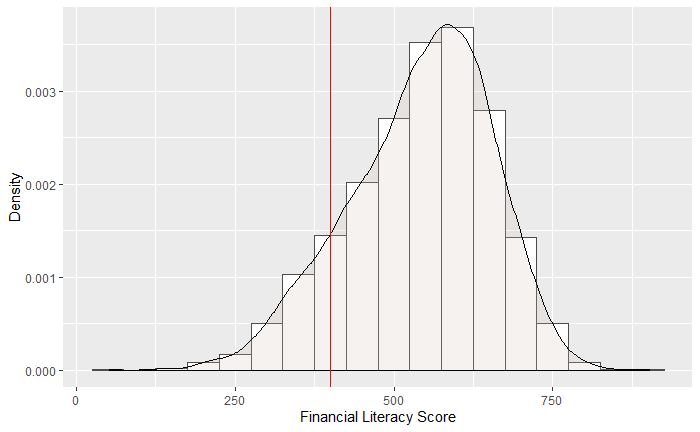}
    }
    \caption{Histogram of the Outcome Variable (PISA Financial Literacy Score). The red line indicates the threshold of the baseline level of proficiency in financial literacy. The OECD suggests that students above this threshold of 400 points have financial literacy levels that are sufficient to participate in society \citep{OECD2017a}.}
    \label{fig:outcome}
\end{figure}

\begin{figure}[H]
    \centering
    \resizebox{0.9\textwidth}{!}{
    \includegraphics{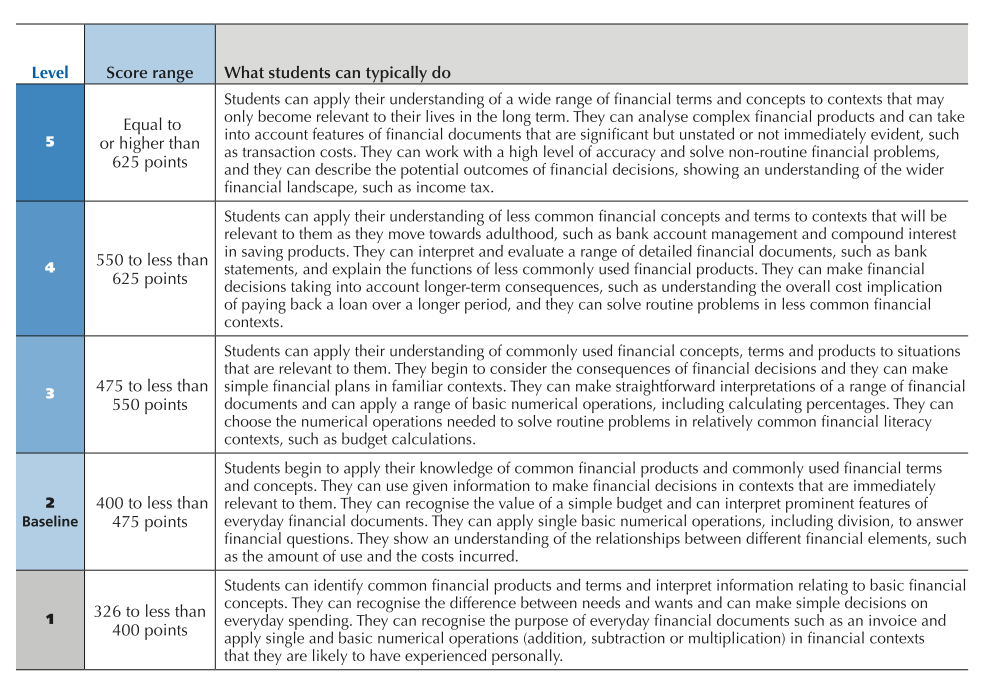}
    }
    \caption{PISA Proficiency Levels for Financial Literacy \citep{OECD2017a}}
    \label{fig:PISAlevels}
\end{figure}

\section{Application Covariates}
This section summarizes the PISA variables used in the analysis, describing their meaning and theoretical relevance for predicting financial literacy.

\hfill \break
\textbf{\textit{Student Characteristics}}

\textbf{ST001D01T – International Grade}:
This variable indicates the student's grade level. Students in PISA are selected according to their age and must be in grade 7 or higher. Grade level captures cumulative exposure to formal schooling, which is directly related to opportunities to acquire numeracy, problem-solving, and financial concepts. Students in higher grades are expected to have had more instructional time relevant to financial literacy.

\textbf{ST004D01T – Gender}: Gender is self-reported by students in the background questionnaire. Gender differences in financial literacy have been documented across countries, often reflecting differences in socialization, confidence with financial decision-making, and exposure to financial tasks. Including gender controls for these systematic differences.

\textbf{AGE – Age}: Age is reported by students and calculated in years. Students in PISA are selected based on their age (rather than grade level) and must be aged between 15 years and 3 months and 16 years and 2 months at the beginning of the testing period. Even within this narrow PISA age range, age captures developmental differences in cognitive maturity and life experience, both of which are relevant for understanding and applying financial knowledge.

\textbf{ISCEDD – Study Track: ISCED Designation}: This variable identifies the ISCED level of the student's educational program. It distinguishes between lower and upper secondary pathways. Study track reflects curricular content and academic intensity, which shape exposure to mathematics, economics-related topics, and applied problem-solving relevant to financial literacy.

\textbf{ISCEDO – Study Track: ISCED Orientation}: 
ISCED orientation differentiates between general, pre-vocational, and vocational programs. It is reported by schools based on program characteristics. Orientation matters because vocational and general tracks differ in their emphasis on applied versus academic skills, potentially influencing students’ familiarity with real-world financial contexts.

\textbf{BELANGN – Speaks Belgian Language at Home}: This variable indicates whether the language spoken at home is one of Belgium's official languages (namely, French, Dutch and German). Language spoken at home serves as a proxy for linguistic integration and access to instructional content. Students who do not speak the instructional language at home may face barriers to understanding financial concepts presented in school or assessments.

\hfill \break
\textbf{\textit{Socioeconomic Status}}

\textbf{HEDRES – Educational Resources at Home}: This index measures the availability of educational resources (e.g., desk, computer, books) in the home, based on student responses. It reflects the learning environment outside school. Greater access to educational resources supports skill development and exposure to information relevant to financial literacy.

\textbf{WEALTH – Family Wealth Index (Economic Possessions)}: The wealth index captures material assets and household possessions reported by students. It represents economic capital rather than educational capital. Family wealth matters for financial literacy because wealthier households are more likely to engage in financial planning, banking, and consumption decisions that expose children to financial concepts.

\textbf{ST013Q01TA – Number of Books at Home}: Students report the approximate number of books available at home. This variable is a long-established proxy for cultural capital and cognitive stimulation. A book-rich environment is associated with higher literacy, numeracy, and abstract reasoning skills that underpin financial literacy.

\textbf{IMMIG – Immigration Status}: Immigration status distinguishes native students, second-generation immigrants, and first-generation immigrants, based on student and parental birthplace. Immigration status captures differences in educational trajectories, language proficiency, and familiarity with host-country financial systems, all of which may affect financial literacy.

\textbf{MISCED – Mother's Education (ISCED)}: Mother's highest level of education is reported by students using ISCED categories. Parental education reflects human capital in the household and is strongly linked to children’s academic outcomes and exposure to financially informed decision-making.

\textbf{FISCED – Father's Education (ISCED)}: Analogous to MISCED, this variable captures father's educational attainment. Including both parents' education allows for a more complete measure of family educational background, which influences norms, expectations, and financial socialization.

\textbf{BMMJ1 – Mother's Job (ISEI)}: This variable codes the mother’s occupation using the International Socio-Economic Index (ISEI), derived from student reports. Occupational status captures labor market position and socioeconomic standing, which are associated with financial knowledge and practices transmitted within the family.

\textbf{BFMJ2 – Father's Job (ISEI)}: Father's occupational status, coded using ISEI, complements maternal occupation. Together, parental occupations provide a multidimensional measure of socioeconomic status relevant to financial exposure and resources.

\textbf{EMOSUPS – Parents' Emotional Support}: This index measures students' perceptions of emotional support from parents, based on questionnaire items. Emotional support is linked to motivation, self-efficacy, and engagement with learning tasks, which can indirectly influence the development of financial literacy.

\hfill \break
\textbf{\textit{Achievement and Attitude}}

\textbf{PV1MATH – Plausible Value 1 in Mathematics}: This is one of the plausible values representing students’ mathematics achievement, estimated using item response theory. Mathematics skills are a core component of financial literacy, as financial tasks require numerical reasoning, proportional thinking, and data interpretation.

\textbf{PV1READ – Plausible Value 1 in Reading}: This is one of the plausible values representing reading proficiency. Financial literacy assessments rely heavily on reading comprehension, interpretation of written information, and understanding complex texts such as contracts or financial statements.

\textbf{REPEAT – Grade Repetition}: This variable indicates whether the student has repeated a grade, based on self-report. Grade repetition is a marker of academic difficulty and disrupted educational trajectories, which may be associated with lower financial literacy.

\textbf{OUTHOURS – Out-of-School Study Time per Week}: Students report the number of hours spent studying outside school. This variable captures learning engagement and academic effort beyond formal instruction, which may contribute to skills accumulation relevant to financial literacy.

\textbf{MMINS – Mathematics Learning Time at School}: This measure reflects the number of minutes per week spent in mathematics instruction. Instructional time in mathematics is directly related to the development of quantitative skills required for financial reasoning. 

\textbf{LMINS – Language Learning Time at School}: Language learning time captures weekly instructional minutes in language subjects. Strong language skills facilitate comprehension of financial information and problem descriptions.

\textbf{ANXTEST – Test Anxiety}: This index measures students’ self-reported anxiety related to testing situations. Test anxiety can negatively affect performance, particularly in applied problem-solving contexts like financial literacy assessments.

\textbf{MOTIVAT – Achievement Motivation}: This index reflects students' self-reported motivation to achieve academically. Motivation influences engagement, persistence, and willingness to tackle complex tasks, all of which are important for acquiring and demonstrating financial literacy.

\hfill \break
\textbf{\textit{School Characteristics}}

\textbf{SC001Q01TA – School Community (Location)}: This variable classifies schools by the urbanicity of their geographic location (e.g., rural, town, city), reported by school principals. Schools' location captures contextual differences in resources, labor markets, and exposure to financial institutions.

\textbf{SC048Q01NA – Share of Students With a Different Heritage Language}: This variable measures the proportion of students whose home language differs from the language of instruction. Linguistic composition may affect instructional practices and peer learning environments relevant to financial literacy.

\textbf{SC048Q02NA – Share of Students With Special Needs}: This variable indicates the proportion of students with special educational needs in the school. It reflects the diversity of learning needs and may influence instructional focus and resource allocation.

\textbf{SC048Q03NA – Share of Socioeconomically Disadvantaged Students}: This measure captures school-level socioeconomic composition. Concentrated disadvantage is associated with fewer resources and lower average achievement, which may affect financial literacy outcomes.

\textbf{SCHSIZE – School Size}: School size reflects total student enrollment. Larger schools may offer more curricular options and resources, while smaller schools may provide more individualized attention; both can influence learning outcomes.

\textbf{CLSIZE – Class Size}: This variable measures the average number of students per class. Class size affects instructional dynamics and teacher attention, which can shape skill development relevant to financial literacy.

\textbf{RATCMP1 – Number of Available Computers per Student}: This indicator reflects the availability of computers for educational use. Access to technology supports digital literacy and exposure to online financial information and tools.

\textbf{LEADPD – Teacher Professional Development}: This variable captures the extent of professional development for school leadership or teachers, based on principal reports. Professional development is linked to instructional quality, which indirectly affects student competencies, including financial literacy.

\textbf{SCHAUT – School Autonomy}: School autonomy measures the degree of decision-making authority at the school level. Greater autonomy may allow schools to tailor curricula and instructional approaches that incorporate applied skills such as financial literacy.

\textbf{EDUSHORT – Shortage of Educational Material}: This index reflects principals’ perceptions of shortages in instructional materials. Resource shortages can constrain teaching effectiveness and learning opportunities. 

\textbf{STRATIO – Student–Teacher Ratio}: This variable measures the number of students per teacher. Lower ratios are generally associated with more individualized instruction, which may support the development of complex applied skills like financial literacy.

\begin{figure}[H]
    \centering
    \resizebox{0.9\textwidth}{!}{
    \includegraphics{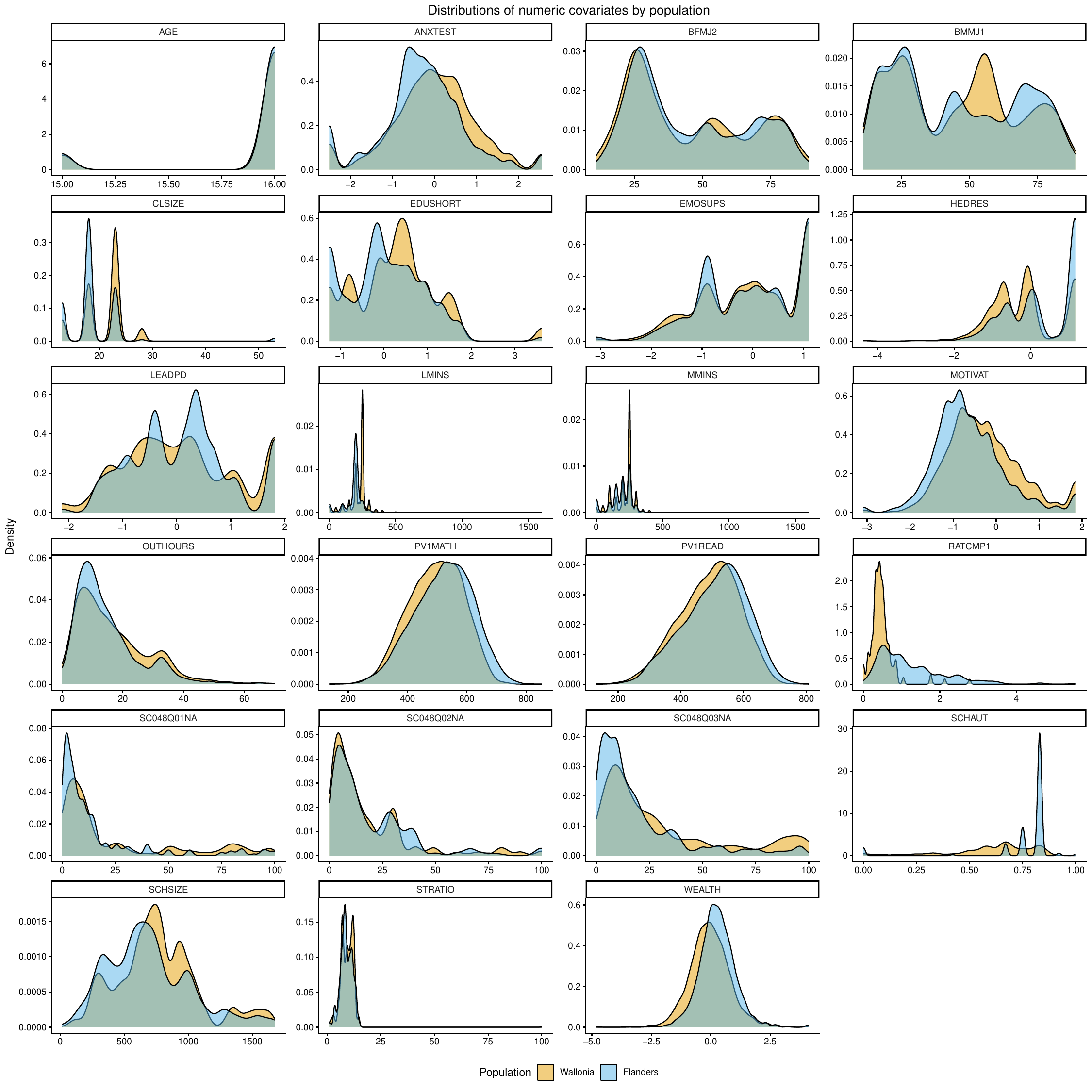}
    }
    \caption{Distribution of numeric covariates, split by region.}
    \label{fig:PISAlevels}
\end{figure}

\begin{figure}[H]
    \centering
    \resizebox{0.9\textwidth}{!}{
    \includegraphics{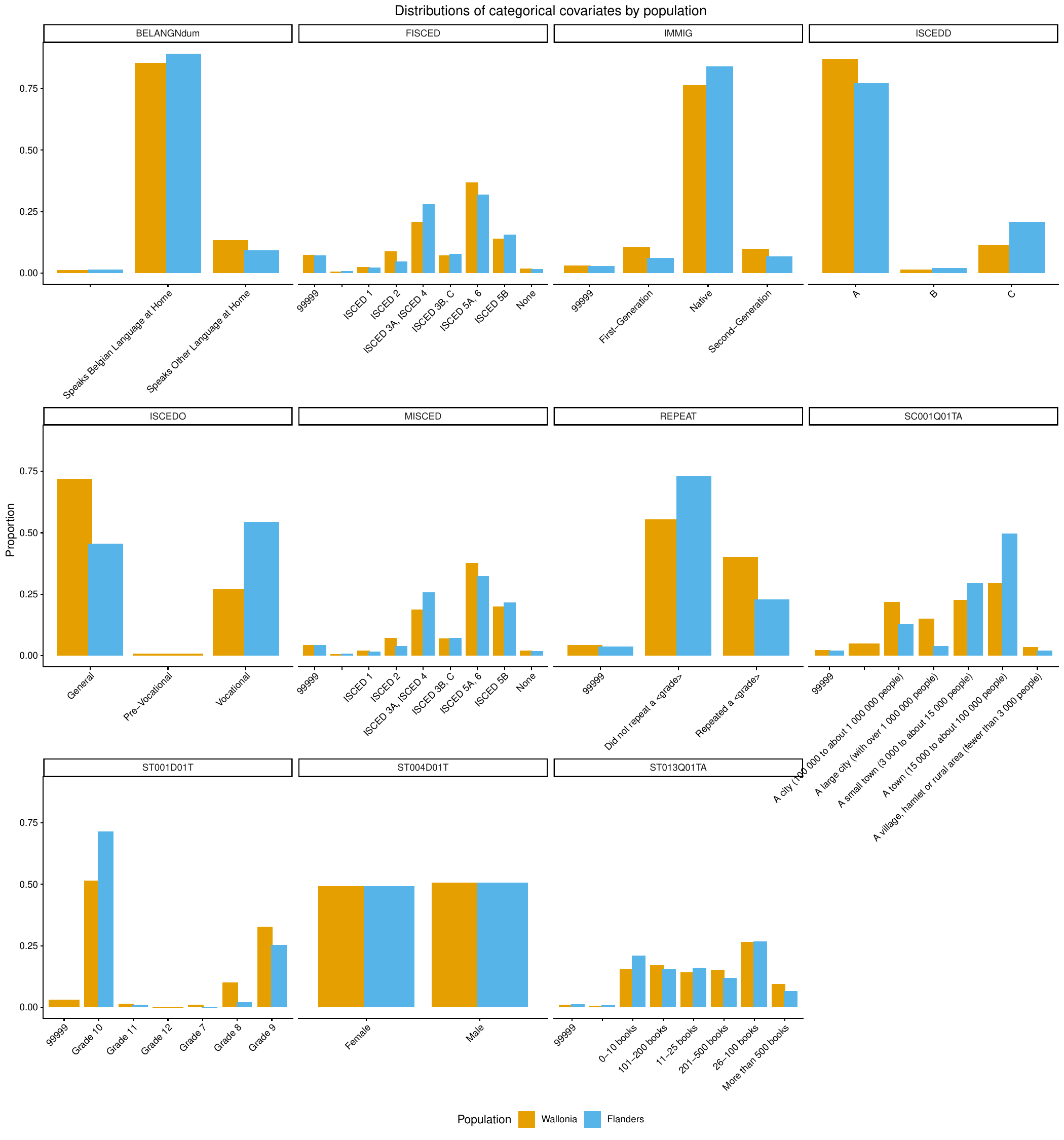}
    }
    \caption{Distribution of categorical covariates, split by region.}
    \label{fig:PISAlevels}
\end{figure}

\doublespacing

\section{Covariate Shift and Overlap} 
\label{sec:covariateshiftappendix}
The BART model has been identified as a model of high predictive performance, and it will therefore be used as a baseline for predictions and modified through reweighting to account for covariate shift. Analyses are run using the \texttt{dbarts} package \citep{dorie2025dbarts}, and we leverage its weighting option to build our proposed model.

\subsection{Overlap Score Estimation} 
As described in Section \ref{sec:covariateshift}, we choose to model our weights following the approach of \cite{steingrimsson2023transporting}, by first obtaining estimates of the overlap scores using machine learning. Approaches such as random forests and BART were considered for the score estimation level, but the simpler logistic regression model was eventually selected as the preferred choice, as it induces more spread in predicted values.

\begin{figure}[H]
    \centering
    \begin{subfigure}[b]{0.32\textwidth}
        \includegraphics[width=\textwidth]{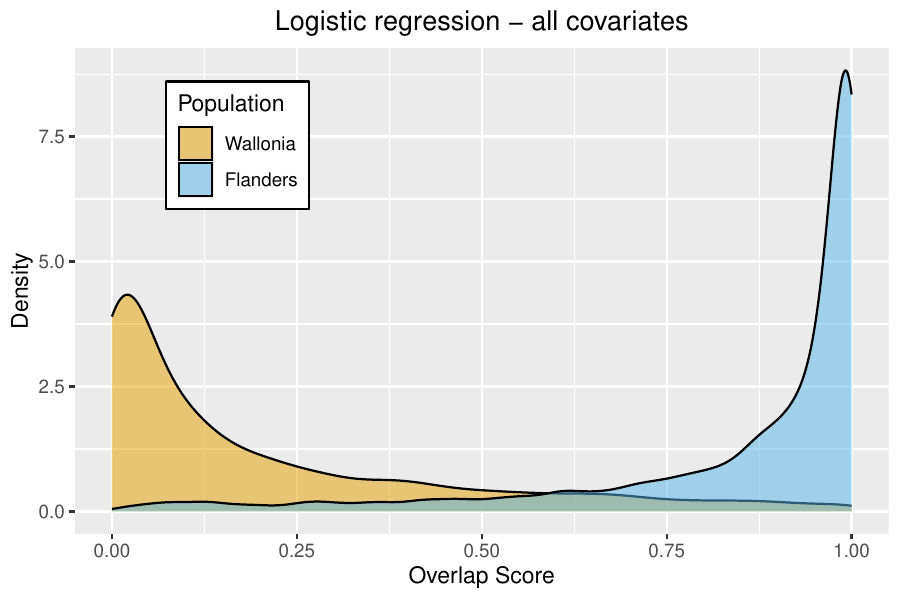}
    \end{subfigure}
    \begin{subfigure}[b]{0.32\textwidth}
        \includegraphics[width=\textwidth]{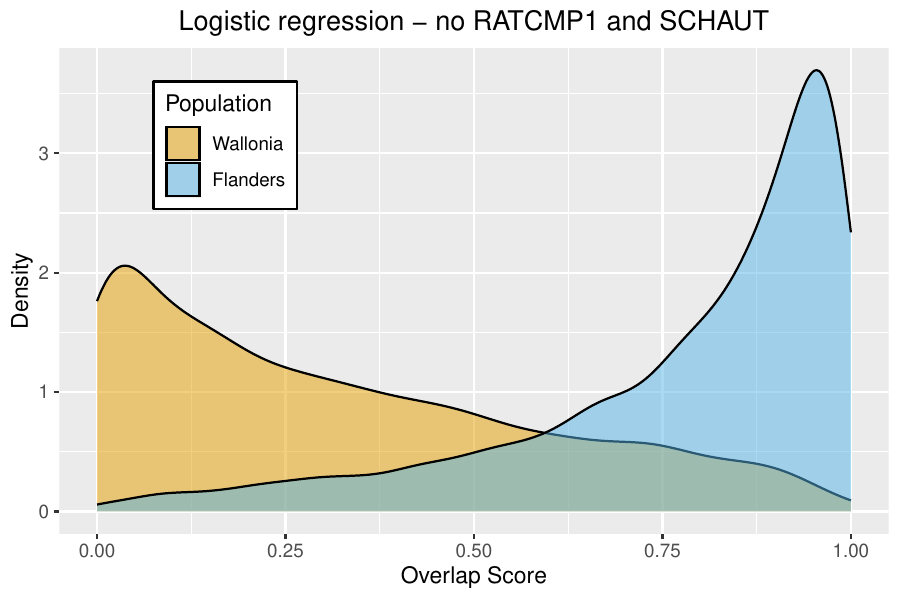}
    \end{subfigure}
    \begin{subfigure}[b]{0.32\textwidth}
        \includegraphics[width=\textwidth]{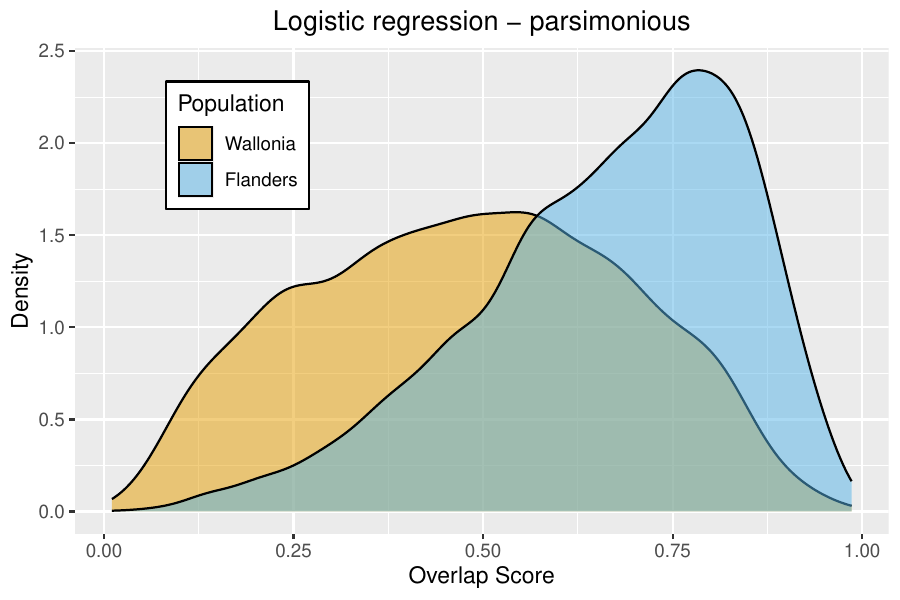}
    \end{subfigure}
    \caption{ Overlap score distributions split by population-- in blue, Flanders, and in yellow, Wallonia. The three panels were obtained from considering different subsets of covariates for estimation using logistic regression. (Left) All covariates, (Middle) all covariates but school autonomy and number of computers per student; and (Right) ten covariates most explanatory of the FLS scores estimates.}
    \label{fig:overlap_dists}
\end{figure}

We conduct a careful analysis to select the covariates included in the estimation of overlap scores. In Figure \ref{fig:overlap_dists}, we compare the distributions of the estimated scores for FLS predictions, considering different covariate set-ups. On the right, we consider a model that includes all available data covariates. We notice the model is very discriminating-- $P(S=1 \mid X)$ is very high for source, i.e., Flanders, observations, and close to zero for the target. This leads to extremely valued weights, which can significantly impact the performance of the adjusted BART algorithm. 

\begin{figure}[H]
    \centering
    \begin{subfigure}[b]{0.48\textwidth}
        \includegraphics[width=\textwidth]{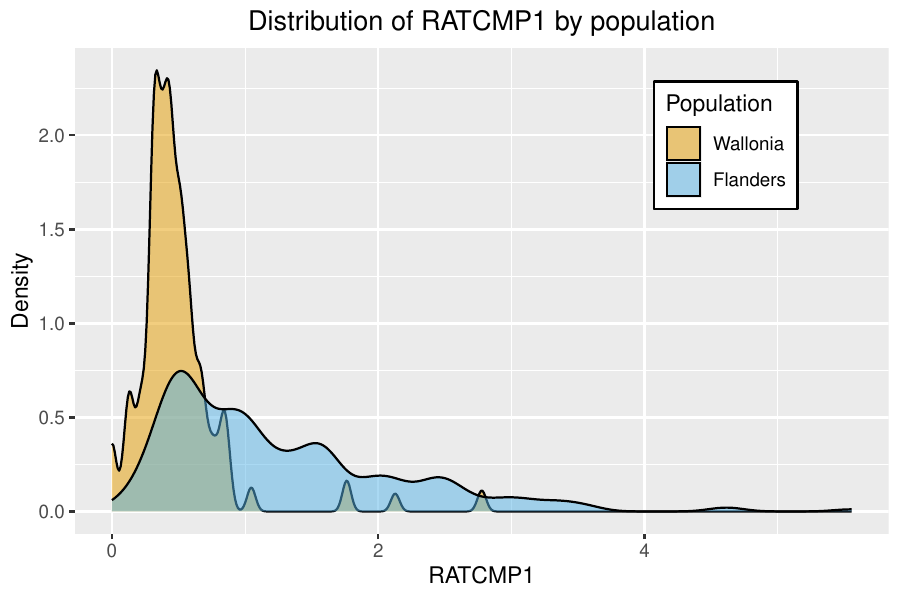}
    \end{subfigure}
    \begin{subfigure}[b]{0.48\textwidth}
        \includegraphics[width=\textwidth]{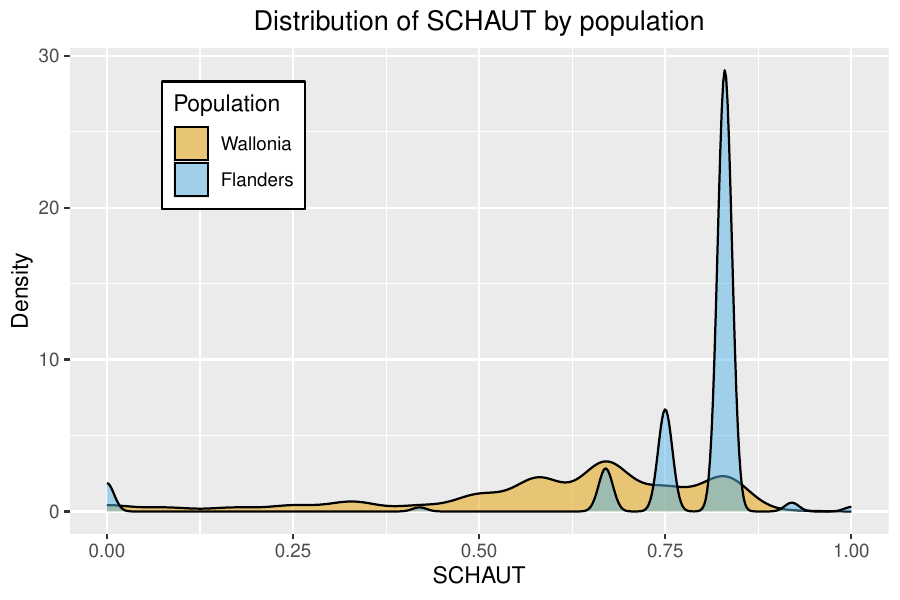}
    \end{subfigure}
    \caption{ Differences in covariate distributions between Flanders (blue) and Wallonia (yellow). (Left)  Distribution of the number of computers per student and (Right) distribution of the school autonomy value.}
    \label{fig:covariate_dists}
\end{figure}

An inspection of variable importance in the score prediction logit regression identifies the variables \texttt{RATCMP1} (number of available computers per student) and \texttt{SCHAUT} (school autonomy) as highly predictive of classification into each region. Specifically, we look at the absolute value of the Wald z-statistic for all predictors in the binomial GLM for population and these two display significantly larger values, indicating high association with the response. Indeed, we see in Figure \ref{fig:covariate_dists} that the distributions display significantly different trends. Those differences are not revealing of a significant lack of overlap, but differentiate the groups clearly. By fitting a linear regression model for the prediction of FLS and investigating the importance of each predictor (defined as the absolute value of the t-statistic for linear models), we find that \texttt{RATCMP1} and \texttt{SCHAUT} have little explanatory power for the outcome. Including them artificially induces poor overlap across regions, without helping to reduce bias in the central prediction model. The central panel of Figure \ref{fig:overlap_dists} shows that the score model trained without those two predictors increases the spread in values. 

Finally, we also consider a much more parsimonious model. Rather than adjusting for shift across all, or most of, the available predictors, we focus on addressing deviations in distributions for covariates most associated with the outcome variable. We include the 10 variables most highly associated with the FLS using importance in a linear regression model. These are identified as : \texttt{PV1READ}, \texttt{PV1MATH}, \texttt{ISCEDD}, \texttt{ANXTEST}, \texttt{HEDRES}, \texttt{FISCED}, \texttt{ST004D01T}, \texttt{MISCED}, \texttt{SCHSIZE} and \texttt{REPEAT}. Under this model, we see a symmetric spread of overlap scores for Wallonia. A peak at higher values still appears for Flanders, but the tails are much wider than under the two other setups.

\subsection{Additional Predictions for Proxy Outcomes}

To evaluate the performance of the considered modeling frameworks, we consider a proxy analysis focusing on the math and reading scores. This study is constructed to evaluate the transportability of predictions-- since the outcome is observed in both populations, errors can be obtained for both source and target data. To obtain estimates on data unseen by the training model, both populations are split into 10 folds for cross-validation. 

\begin{centering}
    \begin{table}[H]
\centering
\small
\caption{Summary of the performance of BART for math and reading scores in Wallonia.}
\label{table:MATHREAD_target_full}
\begin{tabular}{lccc}
\toprule
\textbf{Model Setting} & \textit{RMSE} & \textit{MAE} &  $R^2$ \\
\midrule
\textit{Math Score Outcome} & & & \\
\quad Unweighted     &    51.642    &  41.112      &    0.674   \\
\quad Weighted (1)      &    66.600    &    53.130  &  0.456     \\
\quad Weighted (2)      &    60.130   &    47.801    &  0.556     \\
\quad Weighted (3)      &    52.424    &   41.797     &   0..64    \\
\quad Target-trained &      43.437  &    34.658    &     0.769  \\
\addlinespace
\textit{Reading Score Outcome} & & & \\
\quad Unweighted     & 52.131       &   41.404     &    0.872   \\
\quad Weighted (1)      &  68.457    &   54.699     &     0.632  \\
\quad Weighted (2)      &      61.989  &   49.211     &   0.723    \\
\quad Weighted (3)      &     57.866  &     46.083   &    0.780   \\
\quad Target-trained &    47.550    &  37.746      &  0.906     \\
\bottomrule
\end{tabular}
\end{table}
\end{centering}

Table \ref{table:MATHREAD_target_full} reports the performance results for the prediction of math and reading scores for students in Wallonia using different models trained on the Flanders data. We note that Weighted (1), Weighted (2) and Weighted (3) refer respectively to the overlap modeling using all covariates, all covariates but \texttt{RATCMP1} and \texttt{SCHAUT}, and ten covariates most associated with the relevant outcome. We notice that both for math scores and reading scores, the parsimonious model outperforms the two other weighted models. All results presented in the main text are therefore obtained using the parsimonious overlap model.

\begin{figure}[H]
    \centering
    \begin{subfigure}[b]{0.48\textwidth}
        \includegraphics[width=\textwidth]{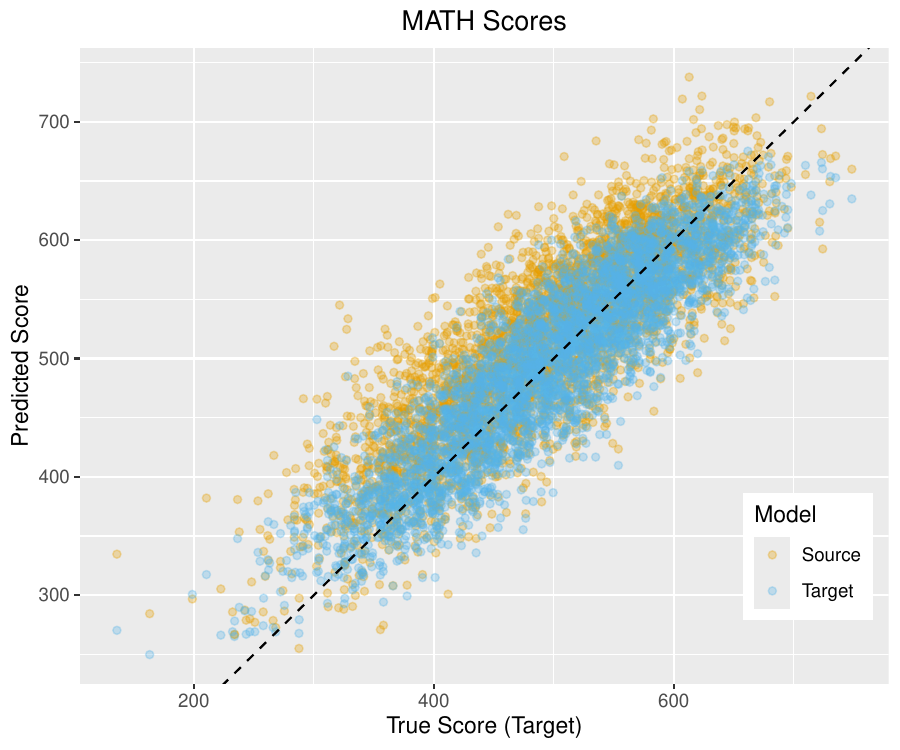}
    \end{subfigure}
    \begin{subfigure}[b]{0.48\textwidth}
        \includegraphics[width=\textwidth]{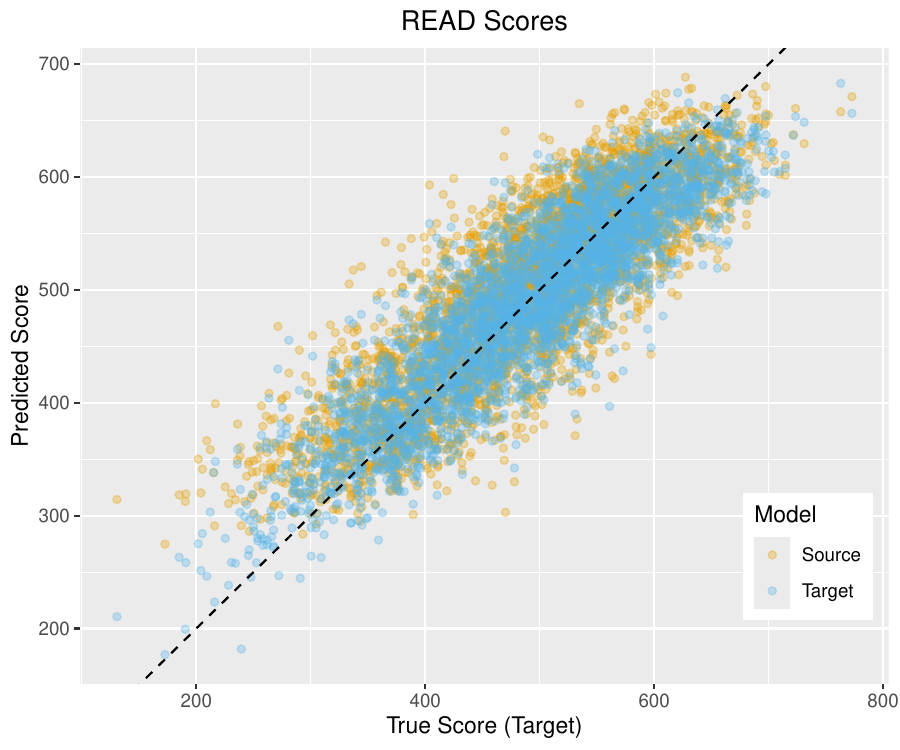}
    \end{subfigure}
    \caption{ (Left) Math scores predictions for Wallonia students using a source-trained (yellow) and target-trained (blue) model. (Right) Reading scores predictions for Wallonia students using a source-trained (yellow) and target-trained (blue) model. The dotted line corresponds to the identity, that is, the points for which the prediction coincides with the true value.}
    \label{fig:math_read_preds}
\end{figure}

In addition to the performance metrics reported, we explore the differences in predictive ability of Wallonia math and reading scores between an (unweighted) BART model trained on the source population and a model trained on the target population. We illustrate individual predictions in \ref{fig:math_read_preds}, where predictions are plotted against the truth, and colored by the model used to make predictions. For the math scores, we find that the predictions obtained using a source model have a tendency to overpredict value compared to the target-trained BART. The differences between the two models are less significant for the reading scores. For both outcomes, we see that the models tend to overestimate predictions of low scores and underestimate those of high scores. Overall, these two plots provide evidence that a model trained on Flanders provides reasonably accurate predictions of the math and reading scores. This serves as justification for using BART to obtain estimates of the missing FLS in Wallonia.

\subsection{Multivariate Overlap}

The results presented in this paper showcase that covariate shift adjustment using a reweighting scheme does not appear to help for the data analysis task at hand. As highlighted by \cite{hooker2004diagnostics}, extrapolation can create unrealistic predictions, even in reasonable points of the predictor space. Hence, we argue that it is critical to get a clear sense of how much the predictive model at hand is relying on extrapolation. We have provided evidence that the strongly discriminating overlap scores are not truly representative of poor covariate overlap, and illustrated the transportability of predictions obtained using the Flanders observations to the Wallonia dataset. Given the relatively large dimensionality of the covariate space, it is, however, difficult to clearly gauge discrepancies in their distributions between regions. We therefore represent the data using a low-dimensional representation obtained through Uniform Manifold Approximation and Projection (UMAP) \citep{mcinnes2018umap}.

\begin{figure}[H]
    \centering
    \begin{subfigure}[b]{0.48\textwidth}
        \includegraphics[width=\textwidth]{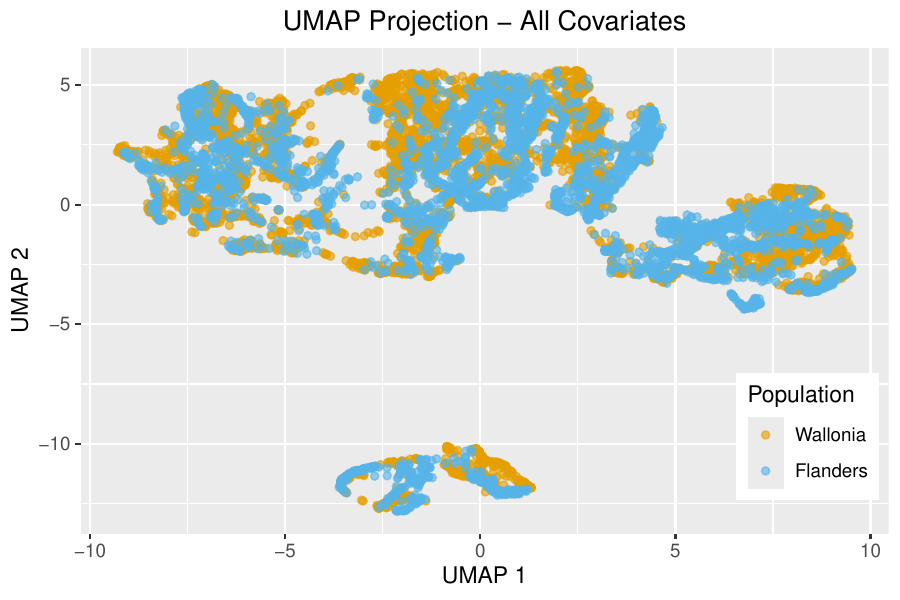}
    \end{subfigure}
    \begin{subfigure}[b]{0.48\textwidth}
        \includegraphics[width=\textwidth]{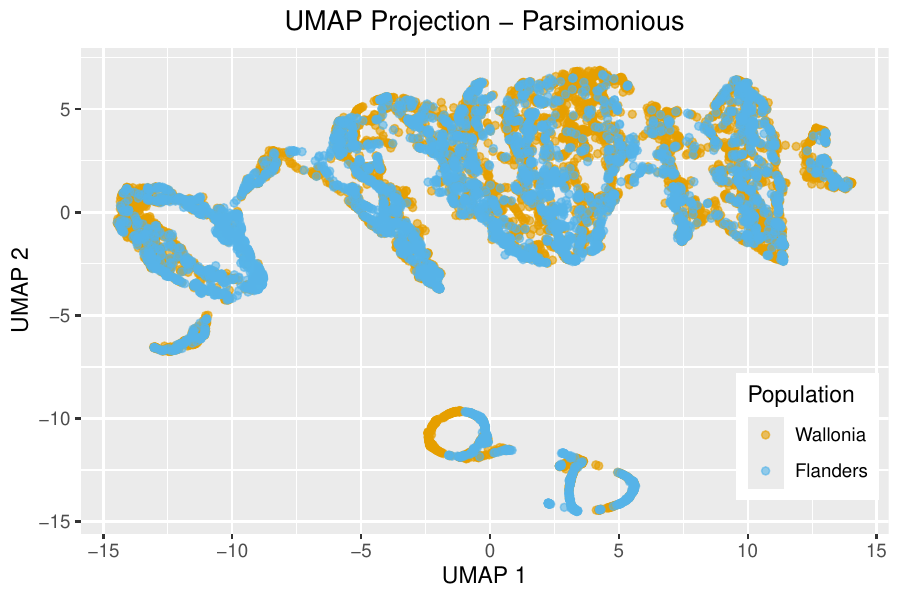}
    \end{subfigure}
    \caption{Two-dimensional UMAP representation, for Flanders (blue) and Wallonia (yellow). The dimensionality reduction is obtained from considering the full covariate set (left) and the parsimonious representation using ten variables most predictive of the FLS outcome (right).}
    \label{fig:umap}
\end{figure}

UMAP is a non-linear dimensionality reduction technique that projects high-dimensional data into 2 or 3 dimensions, helping with visualization and downstream analysis. The data is assumed to lie on a manifold, and the structure is learned by looking at local relationships between points (nearest neighbors). This technique can capture both global and local relationships. We represent the embedding obtained for our covariates in \ref{fig:umap}. The left panel shows the two-dimensional UMAP representation using all covariates; individual observations are colored by the region they belong to. The right panel represents the projection from the parsimonious covariate space considered earlier. In both cases, no clusters arise that differentiate observations from the two regions. The scatter of values along the manifold structure appears similar across Wallonia and Flanders. These provide further evidence for the claim of reasonable overlap in covariates, supporting the transportability of predictions from one region to the other.

\section{Robustness Checks for Covariate Shift Adaptation}\label{appendix:weighting}

To evaluate whether alternative weighting strategies could improve the performance of covariate shift adaptation, we conducted a series of robustness checks using the math scores, for which the target outcome is observed. Weights are estimated using logistic regression, and we consider three covariate sets: (i) all available covariates; (ii) a reduced set excluding \texttt{SCHAUT} and \texttt{RATCMP1} (highly predictive of population membership but uninformative of the outcome); and (iii) only covariates most predictive of the outcome. For each covariate set, we compare four weighting configurations:
\begin{enumerate}
    \item \textbf{Raw weights}: Weights from logistic regression, stabilized as $w_i / \bar{w}$.
    \item \textbf{Winsorized weights}: Weights trimmed at the 5th and 95th percentiles.
    \item \textbf{PCA weights}: Weights estimated from principal components capturing 80\% of covariate variability, computed on pooled source and target data.
    \item \textbf{PCA + Winsorized}: PCA weights trimmed at the 5th and 95th percentiles.
\end{enumerate}

\begin{table}[H]
\centering
\begin{subtable}[t]{0.45\textwidth}
\centering
\caption{Unweighted}
\begin{tabular}{cc}
\toprule
RMSE & MAE \\
\midrule
51.53 & 41.13 \\
\bottomrule
\end{tabular}
\end{subtable}
\hfill
\begin{subtable}[t]{0.45\textwidth}
\centering
\caption{Weighted (All Covariates)}
\begin{tabular}{lcc}
\toprule
Method & RMSE & MAE \\
\midrule
Raw              & 65.62 & 51.62 \\
Winsorized       & 64.13 & 50.94 \\
PCA              & 61.42 & 48.94 \\
PCA + Winsorized & 61.24 & 49.11 \\
\bottomrule
\end{tabular}
\end{subtable}

\vspace{0.8em}

\begin{subtable}[t]{0.45\textwidth}
\centering
\caption{Weighted (Two Removed)}
\begin{tabular}{lcc}
\toprule
Method & RMSE & MAE \\
\midrule
Raw              & 61.09 & 48.43 \\
Winsorized       & 59.90 & 47.45 \\
PCA              & 56.50 & 44.84 \\
PCA + Winsorized & 55.68 & 44.23 \\
\bottomrule
\end{tabular}
\end{subtable}
\hfill
\begin{subtable}[t]{0.45\textwidth}
\centering
\caption{Weighted (Important Covariates)}
\begin{tabular}{lcc}
\toprule
Method & RMSE & MAE \\
\midrule
Raw              & 54.16 & 42.83 \\
Winsorized       & 53.76 & 42.81 \\
PCA              & 53.25 & 42.19 \\
PCA + Winsorized & 53.08 & 42.07 \\
\bottomrule
\end{tabular}
\end{subtable}
\caption{Prediction performance (RMSE and MAE) on target data across weighting strategies. Weights are estimated using source and target covariates; the outcome model is trained on source only.}
\label{tab:weighting_comparison}
\end{table}

Table~\ref{tab:weighting_comparison} presents the results. Even with these adjustments, the unweighted model (RMSE = 51.53, MAE = 41.13) outperforms all weighted variants. Among weighted approaches, the best performance is achieved using outcome-informative covariates with PCA and winsorization (RMSE = 53.08, MAE = 42.07). This pattern suggests that sparser weight models focusing on outcome-relevant covariates are preferred, though they still cannot match the unweighted baseline.

We also examined posterior predictive coverage at the 95\% nominal level. The unweighted model achieves 91\% coverage with an average interval width of 176.2. In contrast, weighted models exhibit substantial undercoverage: raw weights with all covariates yield only 38\% coverage, improving to 72\% with PCA and winsorization. The best coverage among weighted models (89\%) is achieved using outcome-informative covariates with PCA and winsorization, though this still falls short of the unweighted model while producing similar interval widths.

These results reinforce our main findings: in settings with moderate covariate shift, the variance inflation from importance weighting outweighs the theoretical bias reduction, particularly when using flexible nonparametric models such as BART.

\section{Simulation Study: Covariates Distributions in the Three Simulation Scenarios}

\begin{figure}[H]
    \centering

    \begin{subfigure}{0.99\textwidth}
        \centering
        \includegraphics[width=\textwidth]{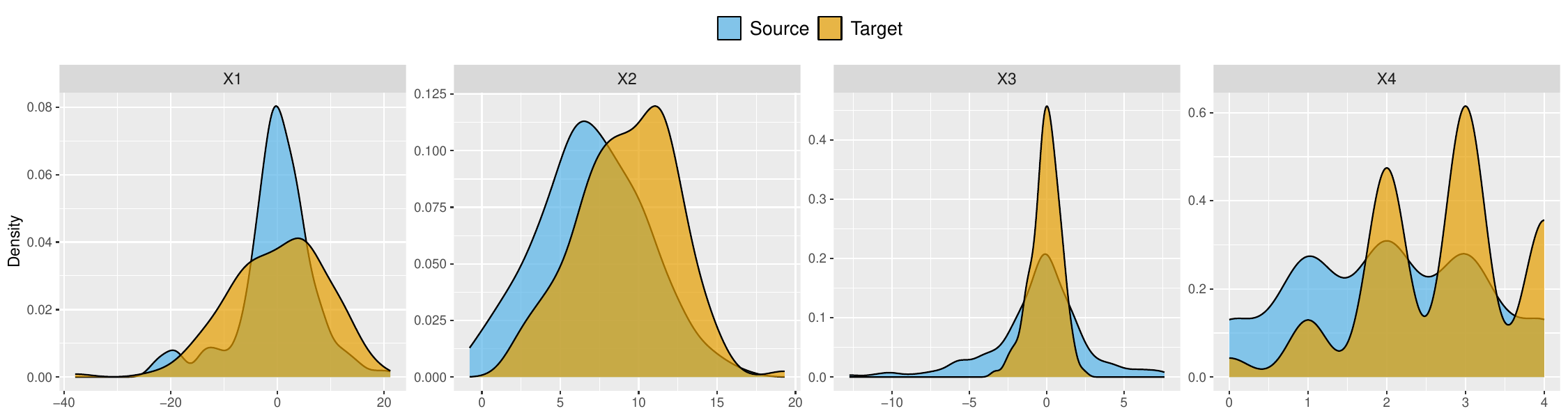}
    \end{subfigure}

    \vspace{0.6em}

    \begin{subfigure}{0.99\textwidth}
        \centering
        \includegraphics[width=\textwidth]{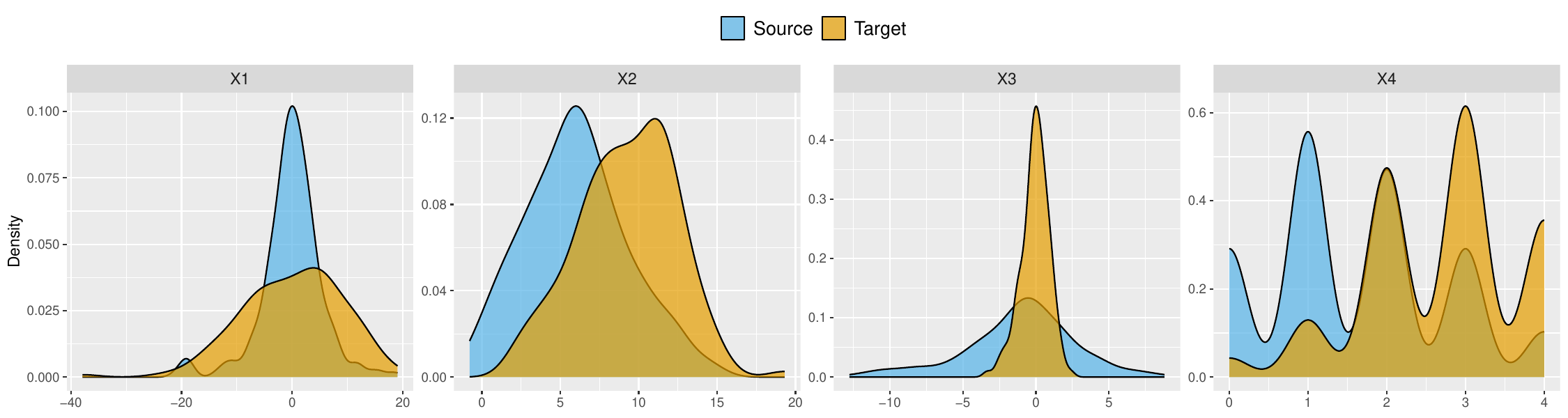}
    \end{subfigure}

    \vspace{0.6em}

    \begin{subfigure}{0.99\textwidth}
        \centering
        \includegraphics[width=\textwidth]{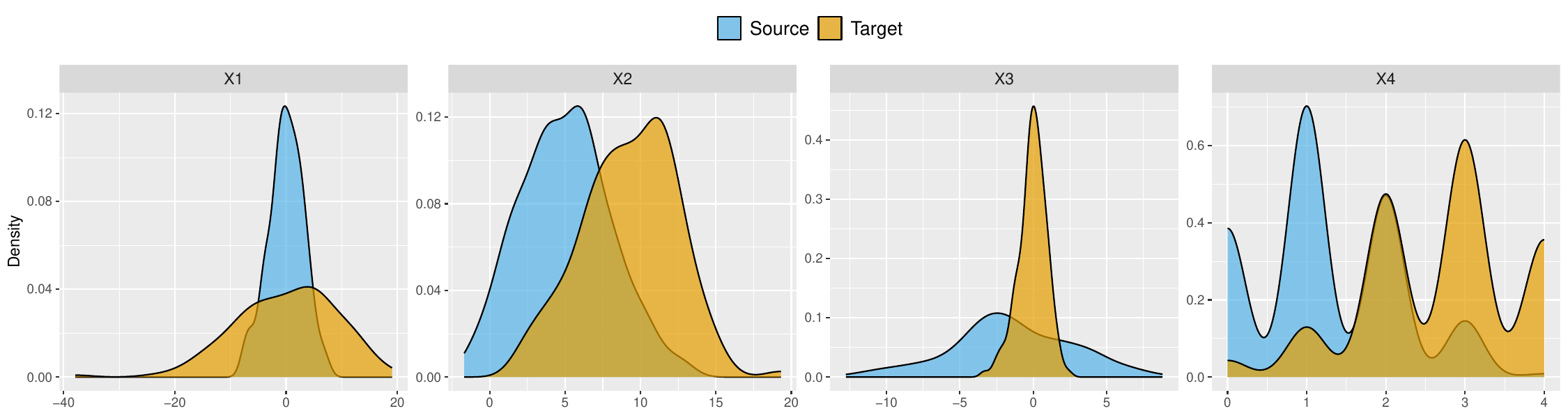}
    \end{subfigure}

    \caption{Marginal covariate distributions for the three simulation scenarios. The top row corresponds to the case where $n_1= n_2 =100$ (Scenario 1), the middle row to the case $n_1 = 150, n_2 =50$ (Scenario 2), and the bottom row to the case where $n_1 = 200, n_2 =0$ (Scenario 3).}
    \label{fig:cov_distributions}
\end{figure}

In Figure~\ref{fig:cov_distributions}, we illustrate the marginal covariate distributions for the three scenarios considered in our simulations. These show the differences in overlap between source and target. The bottom row, corresponding to Scenario 3, where $n_2=0$, has the most apparent differences in covariates distributions, and it has been found to be the scenario in which the reweighting model outperforms the performance of standard BART.

\section{Outlier Plots}

\begin{figure}[H]
    \centering
    \begin{subfigure}[b]{0.48\textwidth}
    \includegraphics[width=\textwidth]{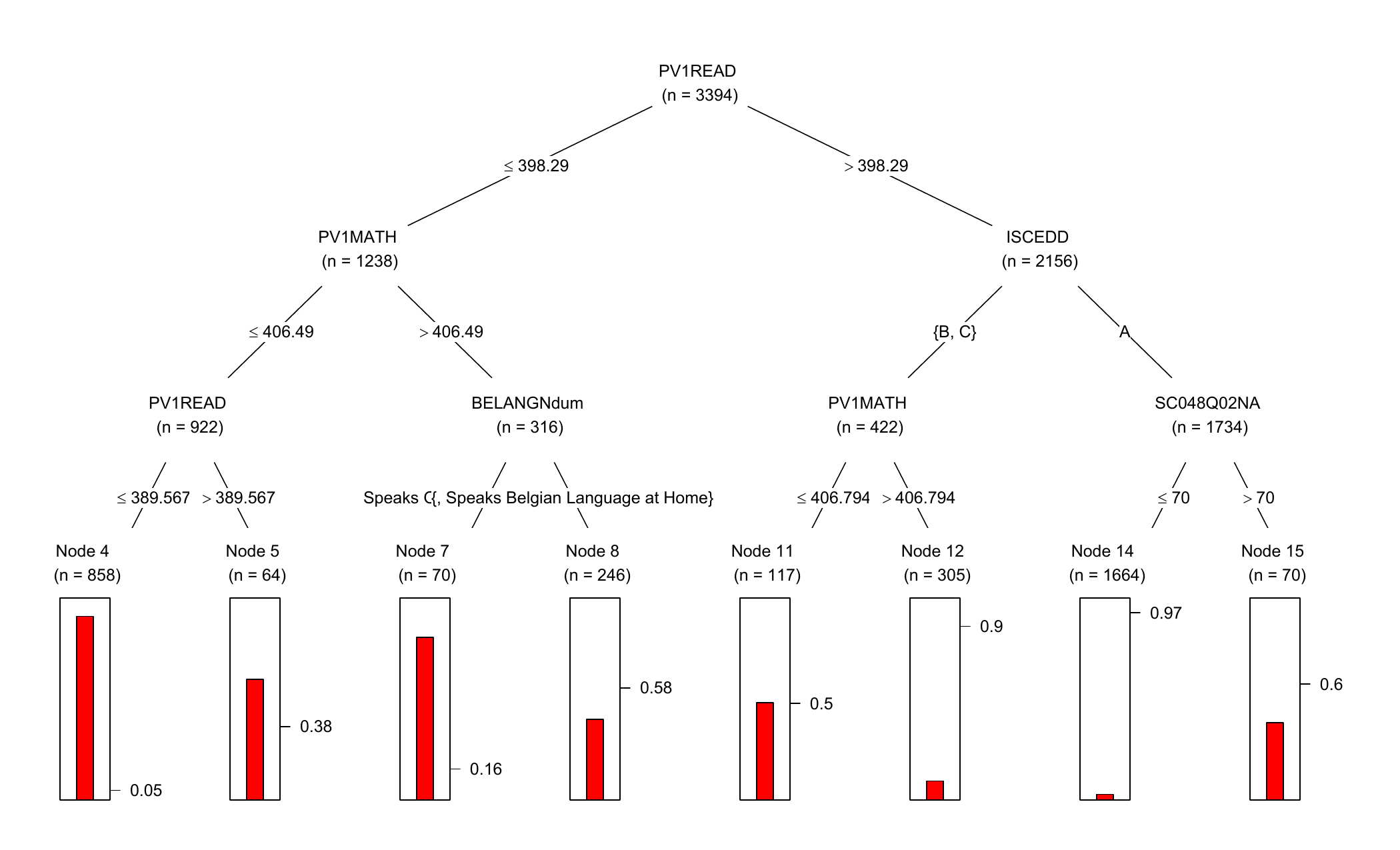}
    \end{subfigure}
    \begin{subfigure}[b]{0.48\textwidth}
    \includegraphics[width=\textwidth]{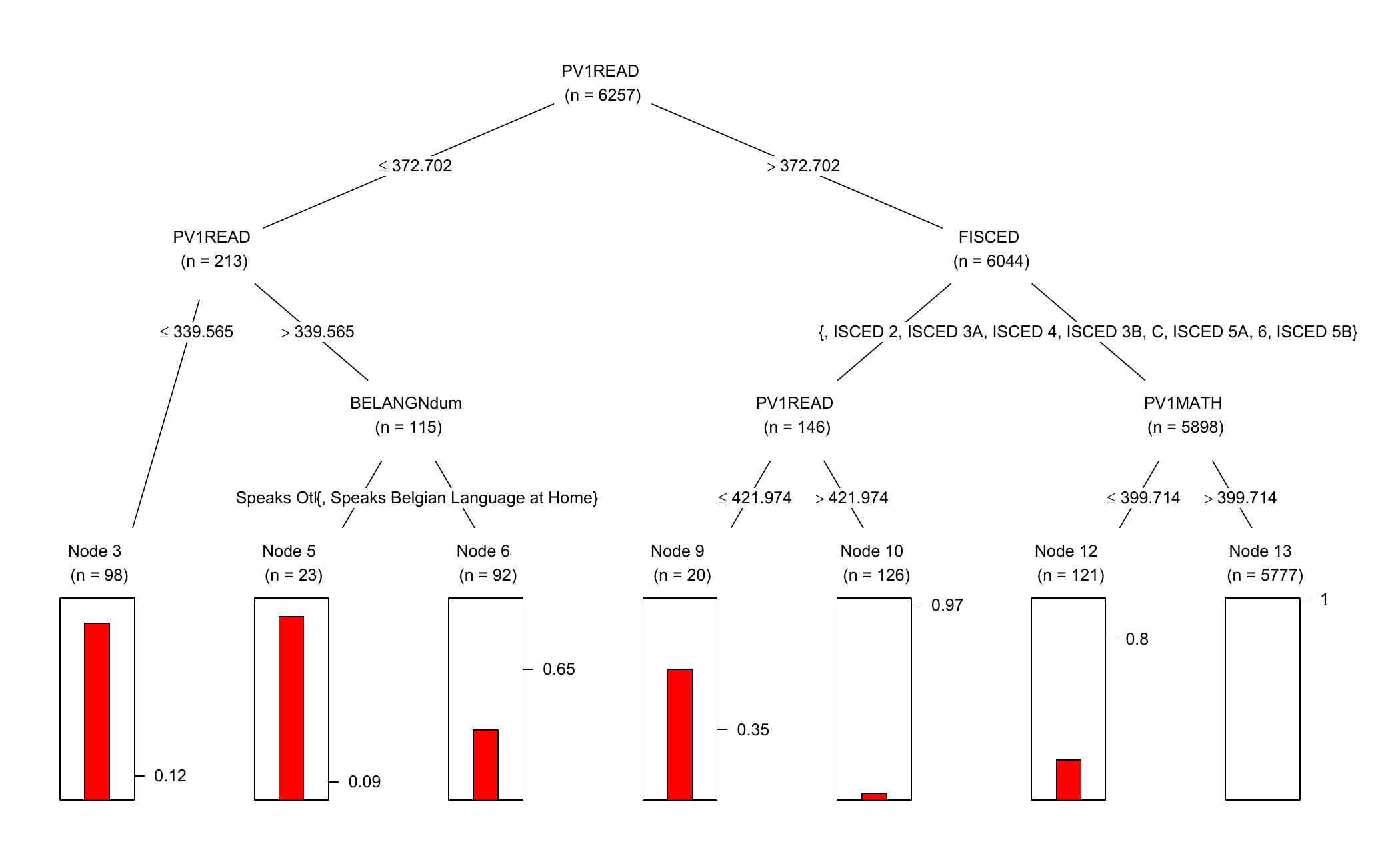}
    \end{subfigure}
    \caption{{\footnotesize(Left) conditional tree for students in grades 7 to 9. (Right) The corresponding tree for students in grades 10 to 12.}}
    \label{fig:ctree_grades1}
\end{figure}

In Figure \ref{fig:ctree_grades1}, we split by grade. We group grades 7-9 (left panel), and 10-12 (right panel). Indeed, in Flanders, after grade 9, certain conditions change (teacher qualification requirements, etc.). Hence, it is interesting to investigate the drivers of heterogeneous effects by grade. In grades 7-9, the largest predictive proportion (95 percent) of students with low FLS are again among those with a lower reading ($\leq398.29$) and math  score ($\leq389.567$). The second largest proportion of students with low FLS (84 percent) is found among non-native speakers with a high math  score ($\>406.49$) but with a low reading score ($\leq398.29$). The same pattern seems to exist in grades 10 to 12. However, in grades 10 to 12, the education of a student's father (\texttt{FISCED}) seems to have a significant influence on FLS. For students with fathers with a higher education background, the predictive probability of low FLS lies between 1 and 20 percent, depending on whether the student has a math  score above or below 399.714.

\begin{figure}[H]
    \centering
    \begin{subfigure}[b]{0.48\textwidth}
    \includegraphics[width=\textwidth]{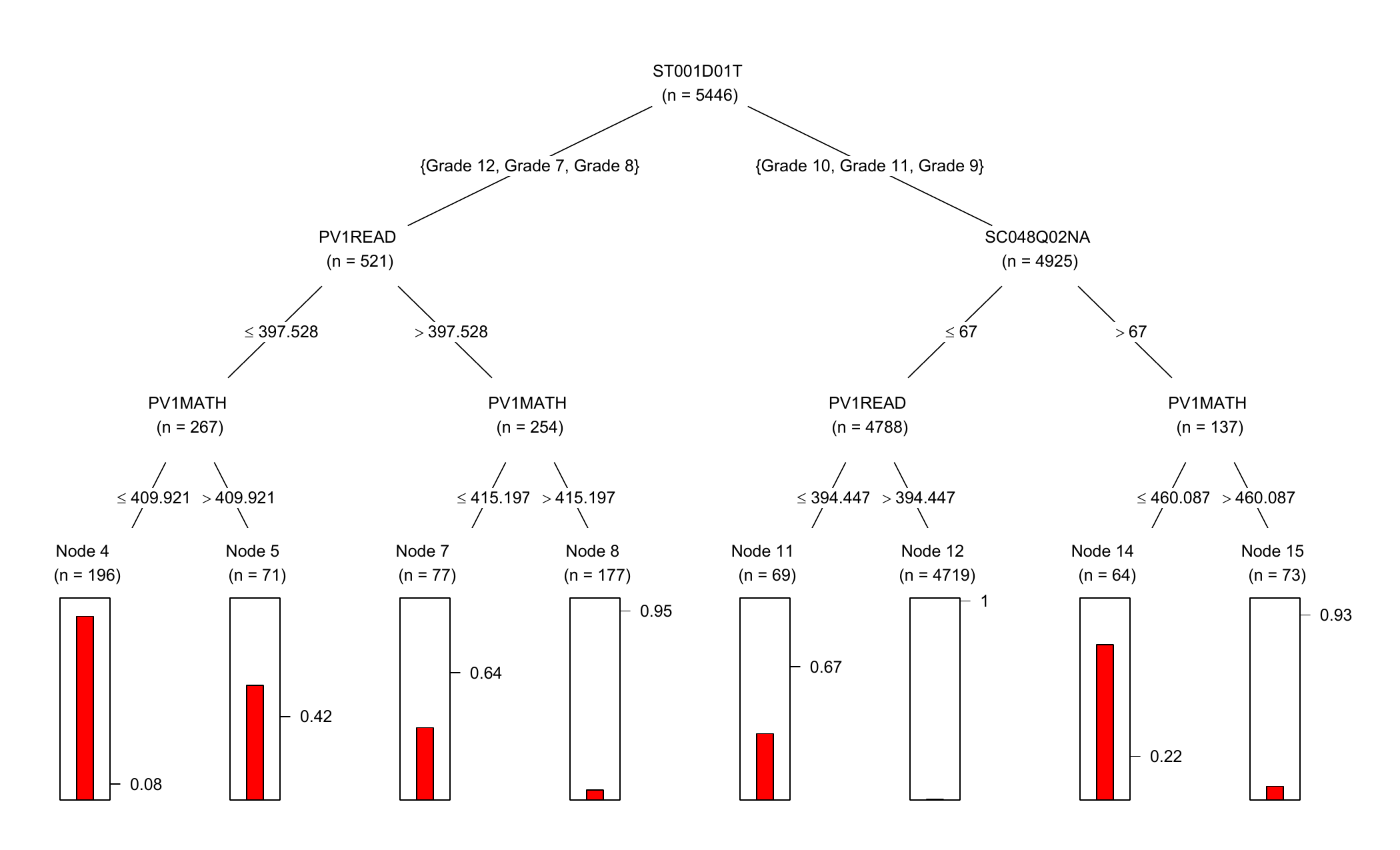}
    \end{subfigure}
    \begin{subfigure}[b]{0.48\textwidth}
    \includegraphics[width=\textwidth]{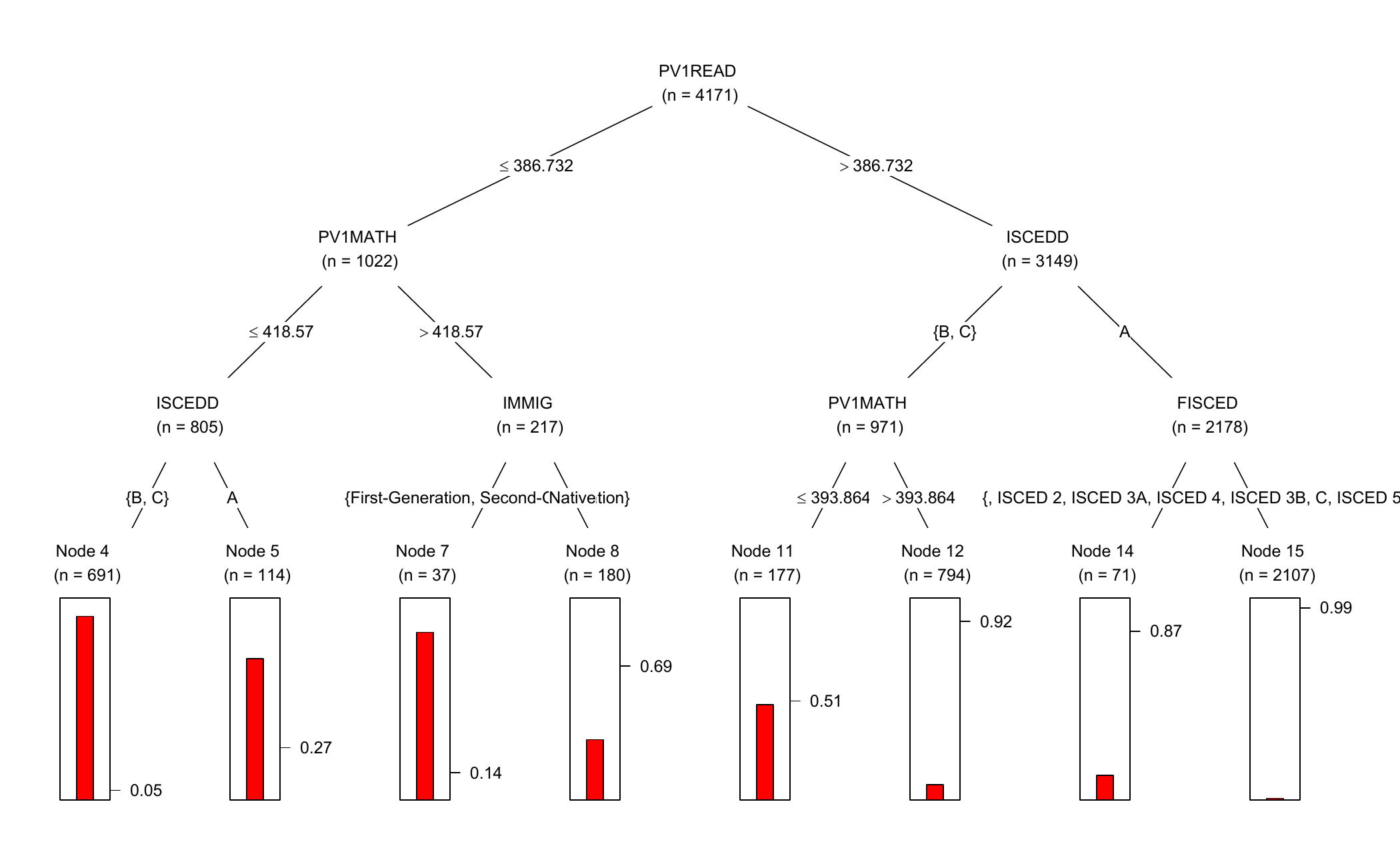}
    \end{subfigure}
    \caption{{\footnotesize(Left) conditional tree for general education. (Right) The corresponding tree for vocational education.}}
    \label{fig:ctree_vocational1}
\end{figure}

In Figure \ref{fig:ctree_vocational1}, we split by track. In the case of general education, students (left panel) from grade 7, 8, and 12 with a low reading ($\leq397.528$) and math score ($\leq409.921$) have, with a predictive probability of 92 percent, the highest chance of having low FLS. The second largest predictive proportion (78 percent) is among students in grades 9, 10, and 11 in schools with more than 67 percent of special needs students (SC0848Q02NA) who have a math score of 460.087 or less. Considering only the students in vocational education (right panel), we observe that the main determinants of students' low financial literacy are the study track, math and reading scores, and whether or not the student speaks Dutch at home.

\section{Changing Outliers Definition} \label{appendix:outliers}
\doublespacing

Below, we depict the results for the conditional tree analysis when we define outliers as observations with predicted values below two absolute deviations from the median. This is a more restrictive definition of outliers. 

\begin{figure}[H]
    \centering
    \includegraphics[width=1\textwidth]{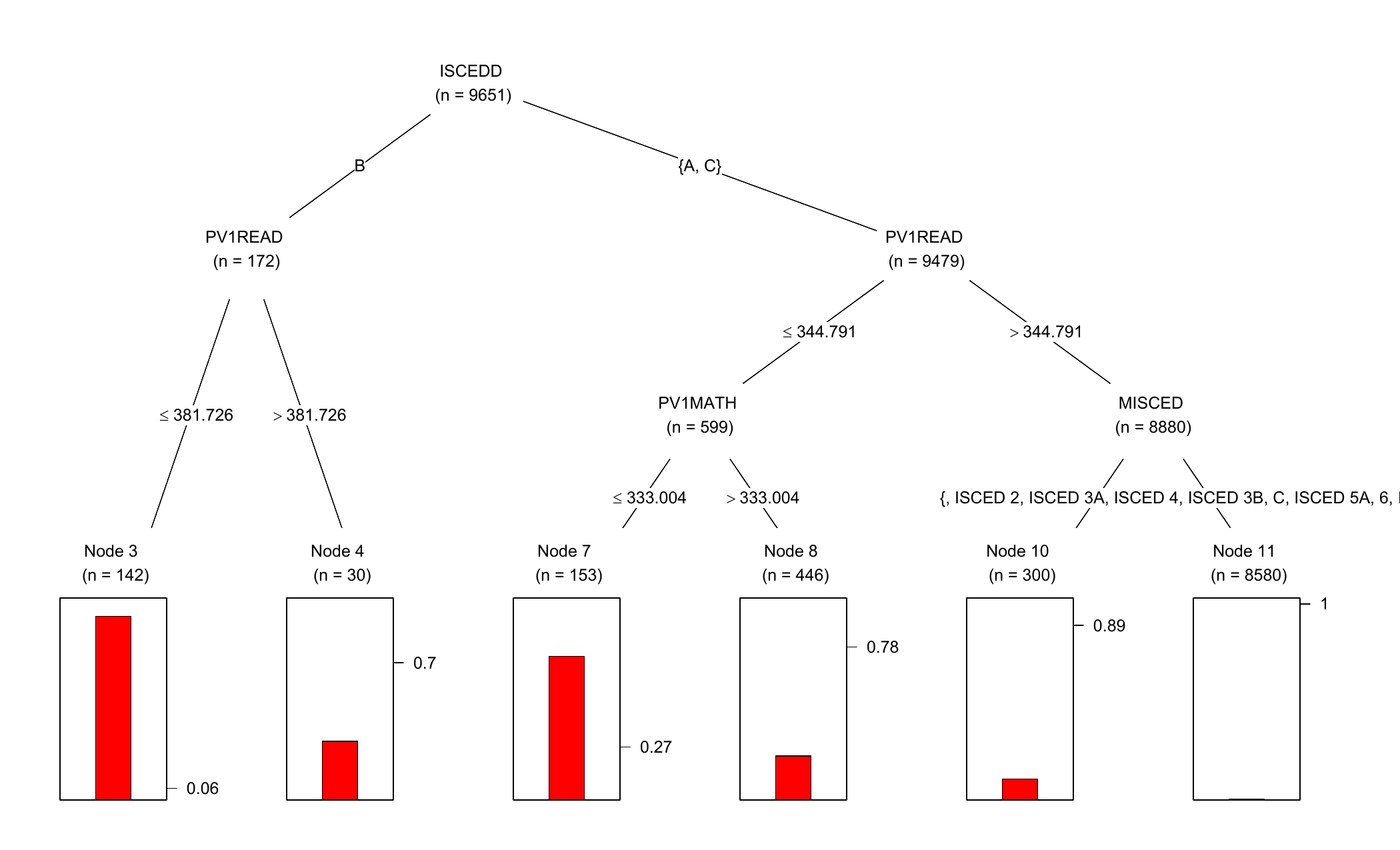}
    \caption{Conditional tree for the entire sample. Within each leaf are depicted in red the histogram of the percentage of units that have a low financial literacy score, and next to it the percentage of units with not-low FLS within the same leaf.}
    \label{fig:ctree_general_mad}
\end{figure}

\begin{figure}[H]
    \centering
    \begin{subfigure}[b]{0.48\textwidth}
    \includegraphics[width=\textwidth]{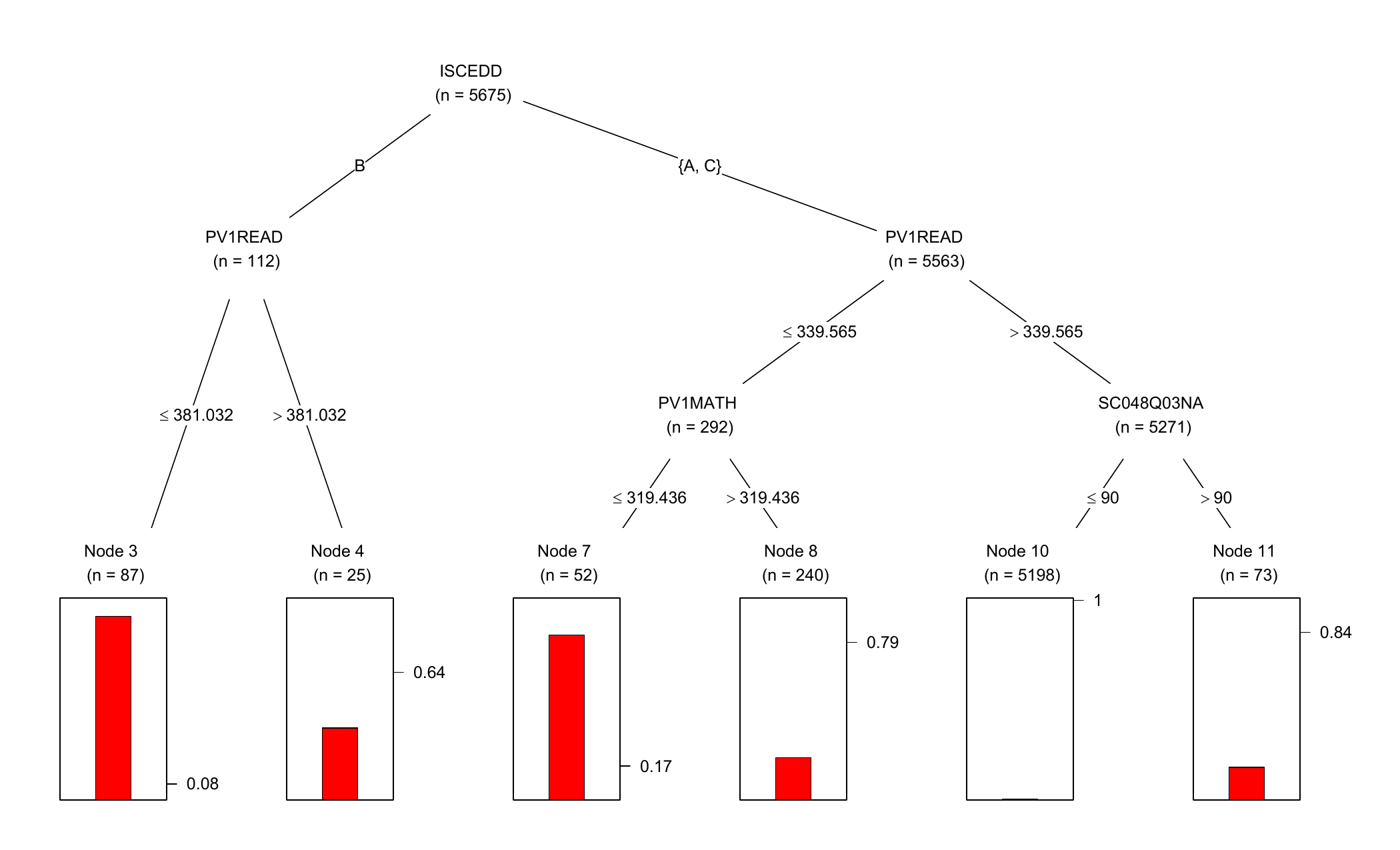}
    \end{subfigure}
    \begin{subfigure}[b]{0.48\textwidth}
    \includegraphics[width=\textwidth]{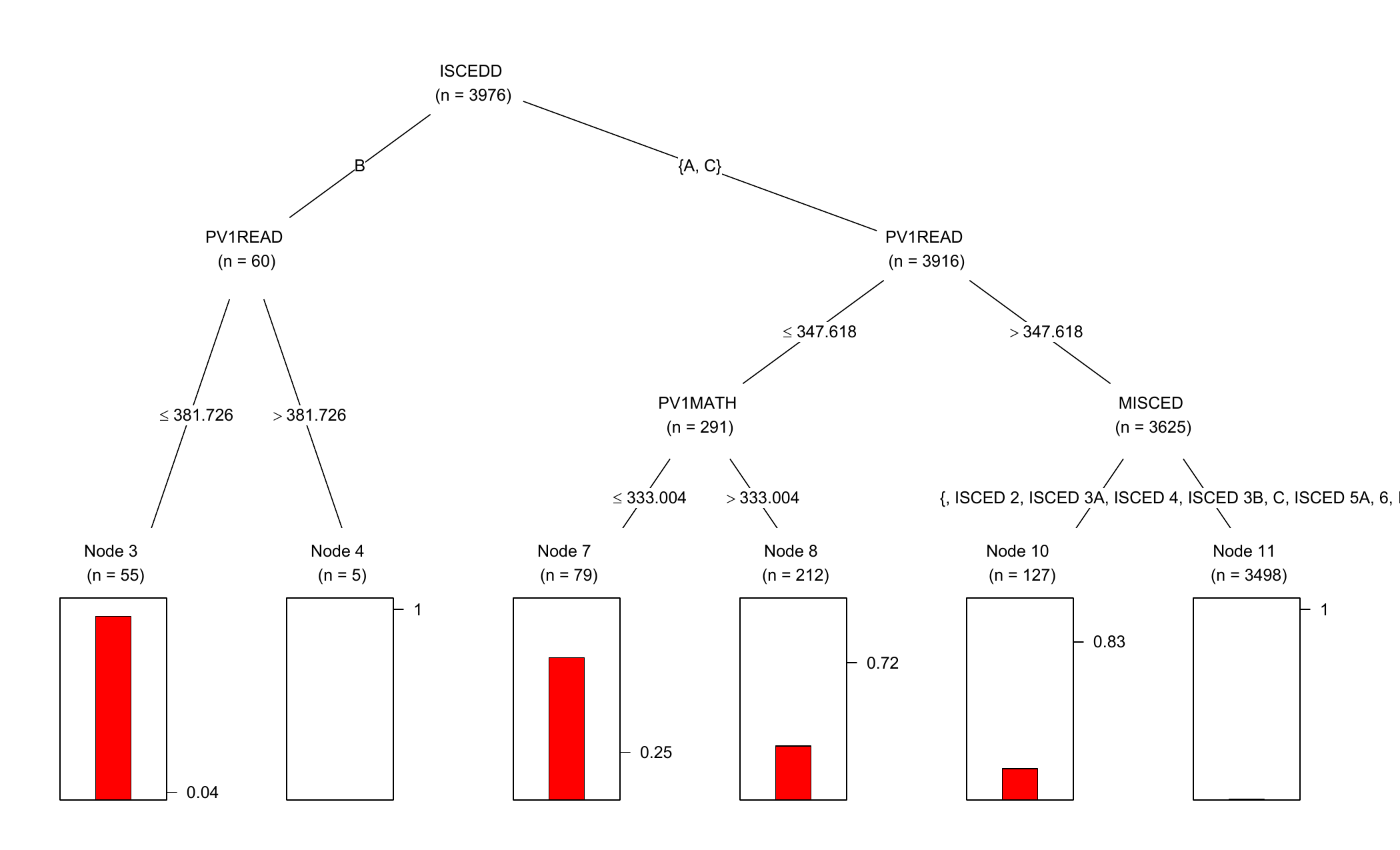}
    \end{subfigure}
    \caption{{\footnotesize(Left) conditional tree for Flanders. (Right) The corresponding tree for Wallonia.}}
    \label{fig:flanders_vs_wallonia}
\end{figure}

\begin{figure}[H]
    \centering
    \begin{subfigure}[b]{0.48\textwidth}
    \includegraphics[width=\textwidth]{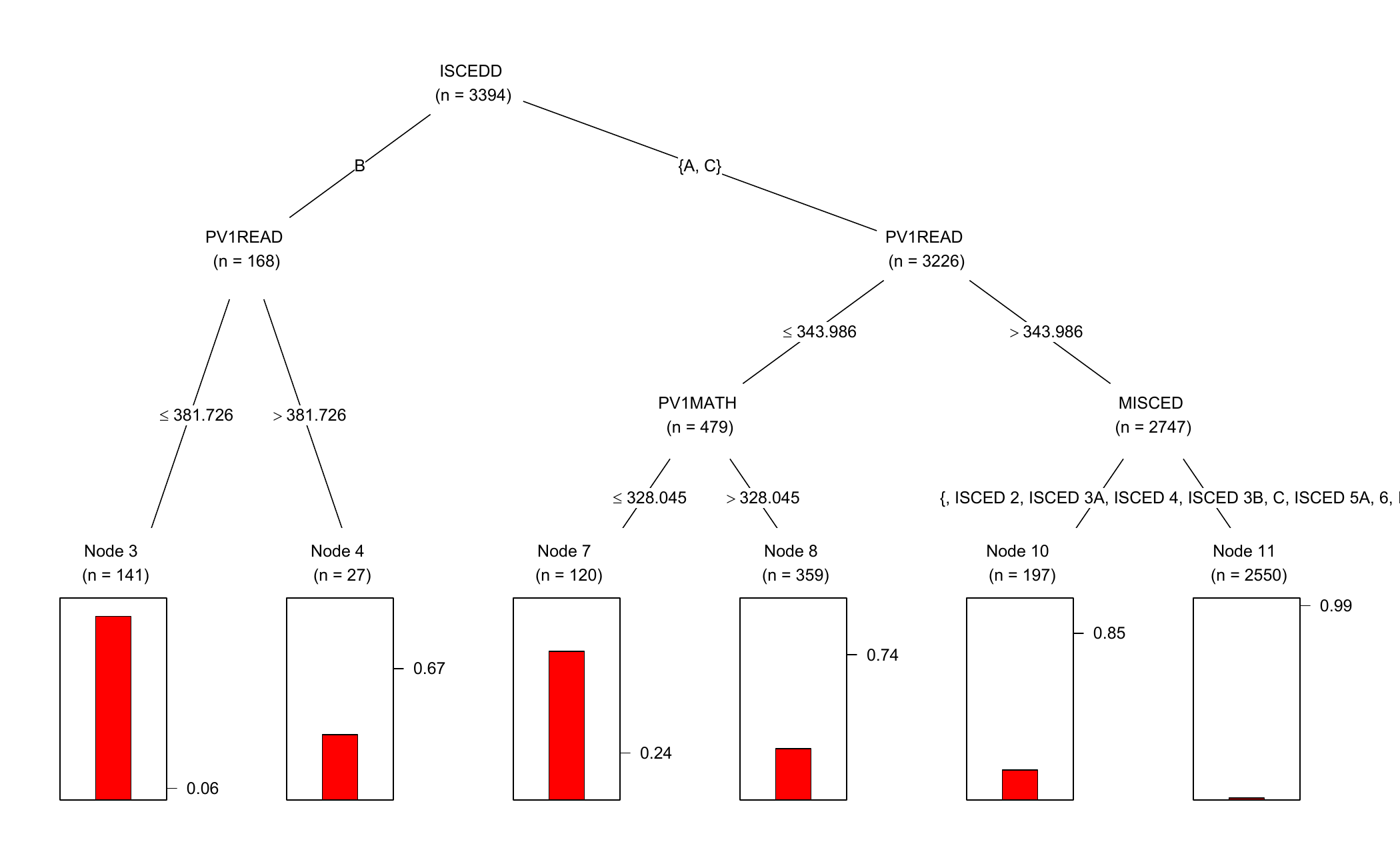}
    \end{subfigure}
    \begin{subfigure}[b]{0.48\textwidth}
    \includegraphics[width=\textwidth]{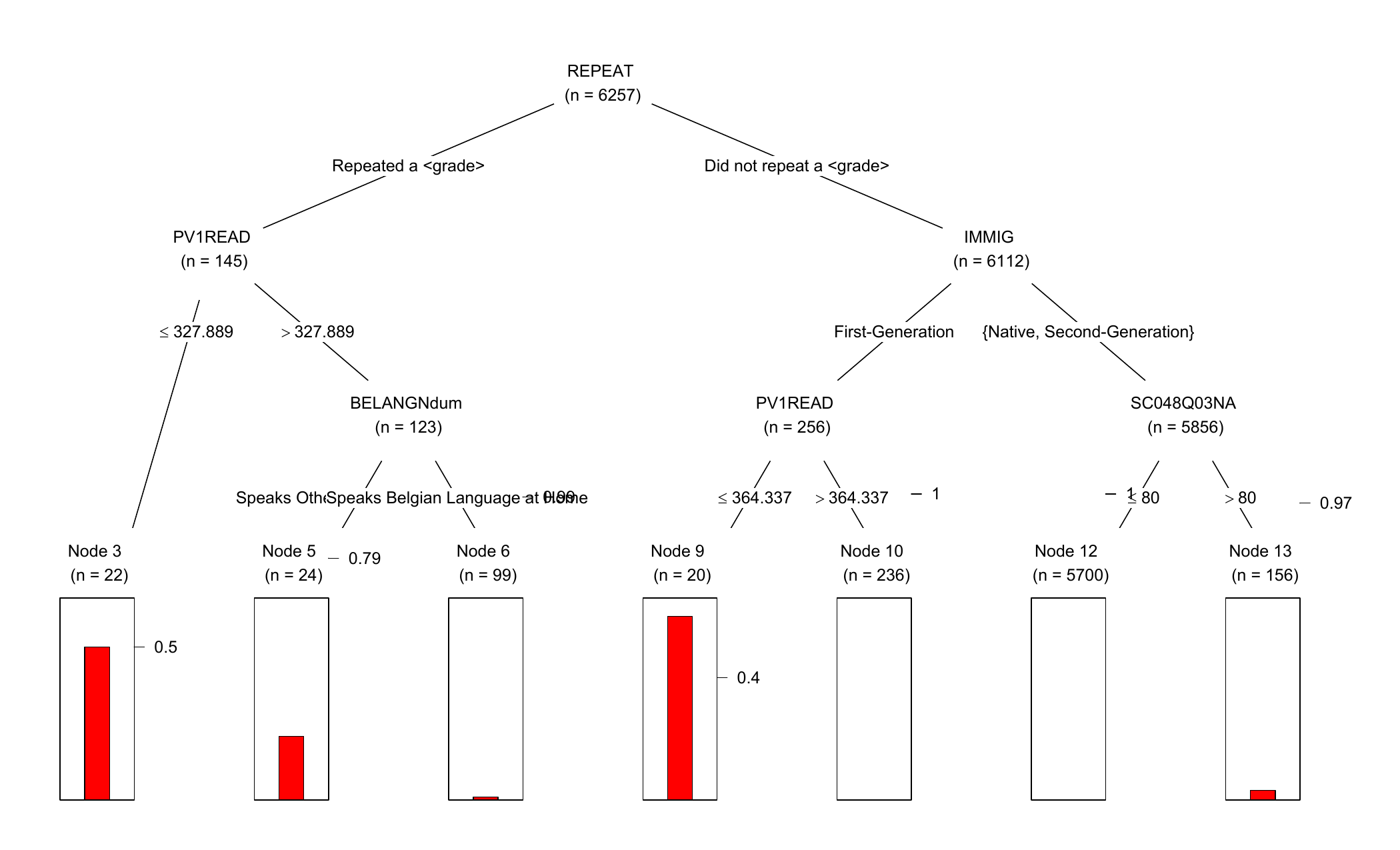}
    \end{subfigure}
    \caption{{\footnotesize(Left) conditional tree for students in grades 7 to 9. (Right) The corresponding tree for students in grades 10 to 12.}}
    \label{fig:ctree_grades}
\end{figure}

\begin{figure}[H]
    \centering
    \begin{subfigure}[b]{0.48\textwidth}
    \includegraphics[width=\textwidth]{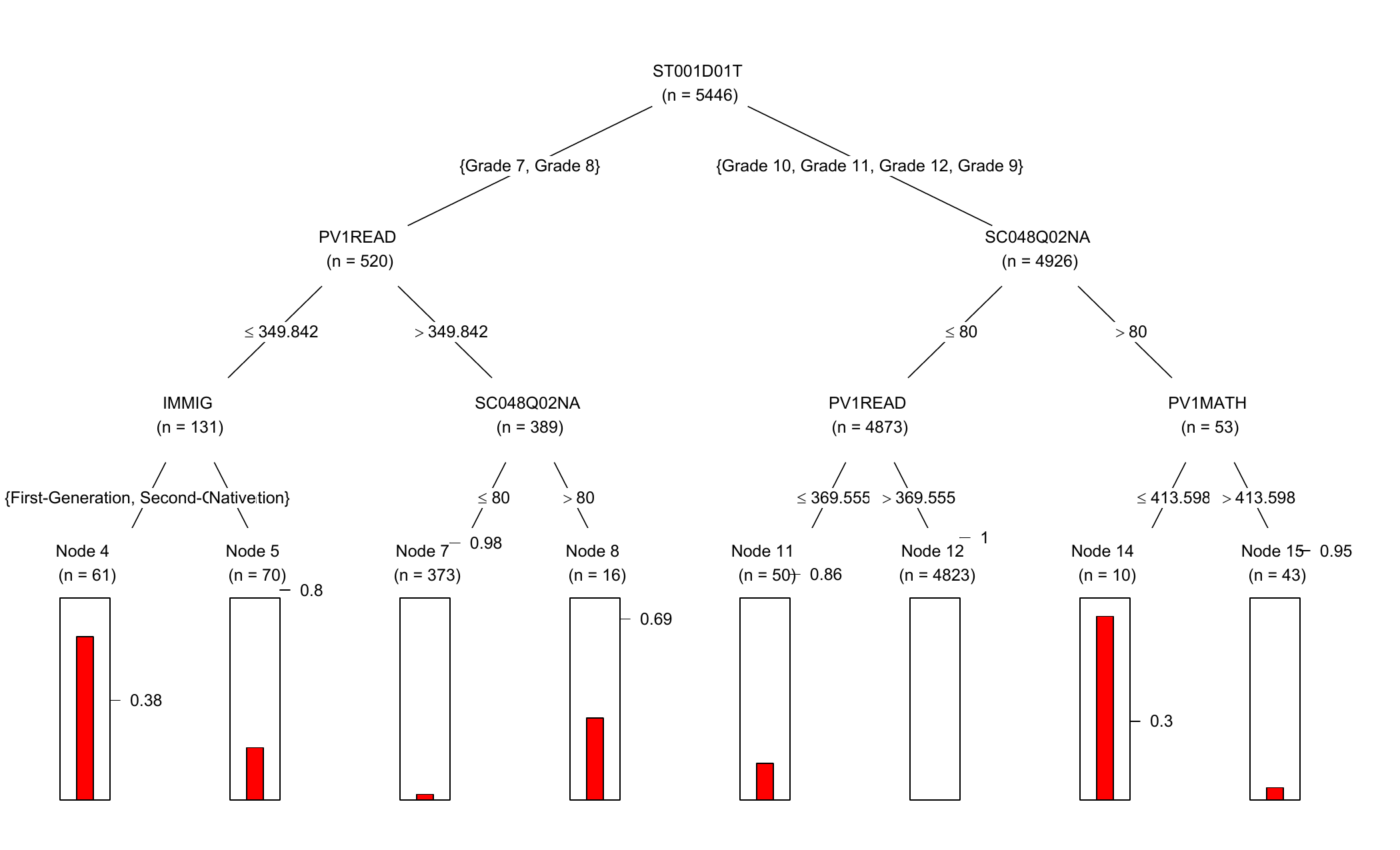}
    \end{subfigure}
    \begin{subfigure}[b]{0.48\textwidth}
    \includegraphics[width=\textwidth]{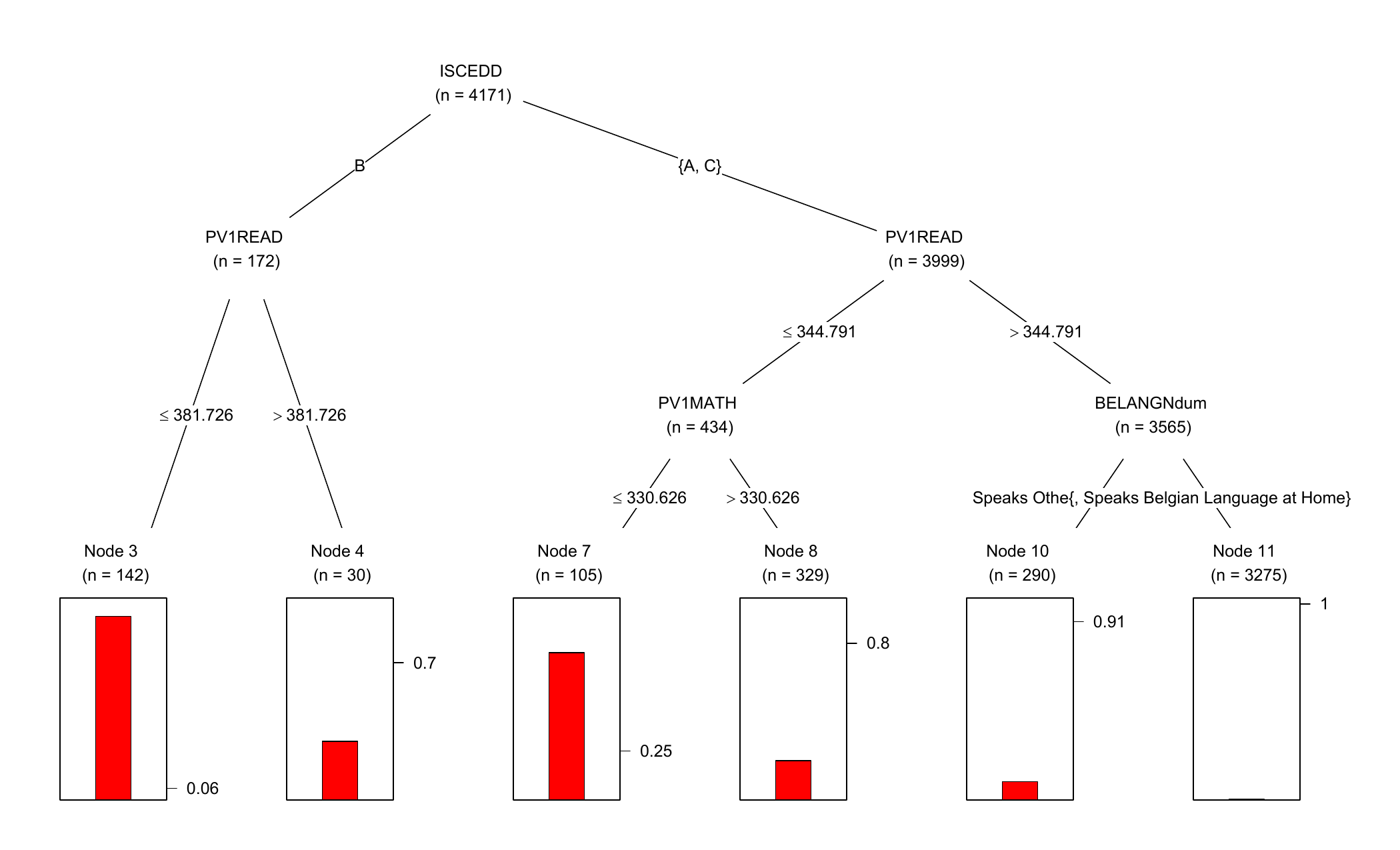}
    \end{subfigure}
    \caption{{\footnotesize(Left) conditional tree for general education. (Right) The corresponding tree for vocational education.}}
    \label{fig:ctree_vocational}
\end{figure}

\end{document}